\documentclass[pre,twocolumn,showpacs,byrevtex]{revtex4b3}

\usepackage{graphicx,epsfig,verbatim,enumerate}
\usepackage{amssymb}
\usepackage{ifthen}

\newcommand{\thnote}[1]{}
\newcommand{\footcite}[1]{~\cite{#1}}

\newcommand{\etal}{\textit{et al.}}
\newcommand{\www}[1]{\url{#1}}
\newcommand{\req}[1]{(\ref{#1})}

\newcommand{\Om}{\Omega}
\newcommand{\om}{\omega}

\newcommand{\ddiff}[1]{\frac{\mbox{d}}{\mbox{d} #1}}

\newcommand{\cprob}[2]{P(#1\,|\,#2)}

\newcommand{\dee}[1]{\mbox{d}#1 \,}
\newcommand{\postdee}[1]{\,\mbox{d}#1}

\newcommand{\avg}[1]{\left\langle#1\right\rangle}
\newcommand{\tavg}[1]{\langle#1\rangle}

\newcommand{\okell}{{l^{\mbox{\scriptsize{(s)}}}}}

\newcommand{\okellsombar}{\bar{l}_{\om}^{\mbox{\scriptsize{\, (s)}}}}

\begin{document}

\title{
Geometry of River Networks I:\\ Scaling, Fluctuations, and Deviations
}

\author{
  \firstname{Peter Sheridan}
  \surname{Dodds}
  }
\thanks{Author to whom correspondence should be addressed}
\email{dodds@segovia.mit.edu}
\homepage{http://segovia.mit.edu/}
\affiliation{Department of 
Mathematics and Department of Earth, 
Atmospheric and Planetary Sciences,
Massachusetts Institute of Technology,
Cambridge, MA 02139.}

\author{
  \firstname{Daniel H.}
  \surname{Rothman}
  }
\email{dan@segovia.mit.edu}
\affiliation{Department  of Earth, 
Atmospheric and Planetary Sciences,
Massachusetts Institute of Technology, 
Cambridge, MA 02139.}

\date{\today}

\begin{abstract}
This article is the first in a series of three papers investigating
the detailed geometry of river networks.  Branching networks are a
universal structure employed in the distribution and collection of
material.  Large-scale river networks mark an important class of
two-dimensional branching networks, being not only of intrinsic
interest but also a pervasive natural phenomenon.  In the description
of river network structure, scaling laws are uniformly observed.
Reported values of scaling exponents vary suggesting that no unique
set of scaling exponents exists.  To improve this current
understanding of scaling in river networks and to provide a fuller
description of branching network structure, here we report a
theoretical and empirical study of fluctuations about and deviations
from scaling.  We examine data for continent-scale river networks such
as the Mississippi and the Amazon and draw inspiration from a simple
model of directed, random networks.  We center our investigations on
the scaling of the length of sub-basin's dominant stream with its
area, a characterization of basin shape known as Hack's law.  
We generalize this relationship to a joint probability density and
provide observations and explanations of deviations from scaling.  
We show that fluctuations about scaling are substantial and grow with
system size.  We find strong deviations from scaling at small scales
which can be explained by the existence of linear network structure.
At intermediate scales, we find slow drifts in exponent values
indicating that scaling is only approximately obeyed and that
universality remains indeterminate.  At large scales, we observe a
breakdown in scaling due to decreasing sample space and correlations
with overall basin shape.  The extent of approximate scaling is
significantly restricted by these deviations and will not be improved
by increases in network resolution.
\end{abstract}


\maketitle

\section{Introduction}
\label{sec:dev.intro}

Networks are intrinsic to a vast number of complex forms
observed in the natural and man-made world.
Networks repeatedly
arise in the distribution and sharing
of information, stresses and materials.
Complex networks give rise to interesting mathematical
and physical properties as observed in
the Internet~\cite{reka99},
the ``small-world'' phenomenon~\cite{watts98},
the cardiovascular system~\cite{zamir99},
force chains in granular media~\cite{coppersmith96},
and the wiring of the brain~\cite{cherniak99}.

Branching, hierarchical geometries make up an
important subclass of all networks.
Our present investigations concern
the paradigmatic example of river networks.
The study of river networks, though
more general in application, is an integral part of
geomorphology, the theory of earth surface processes and form.
Furthermore, river networks are
held to be natural exemplars of allometry, i.e.,
how the dimensions of different parts of a structure scale or grow with
respect to each other~\cite{mandelbrot83,rodriguez-iturbe97,rinaldo98,ball99,dodds2000pa}.  
The shapes of drainage basins, for example, 
are reported to elongate with increasing basin size~\cite{hack57,maritan96a,rigon96}.

At present, there is no
generally accepted theory explaining
the origin of this allometric scaling.
The fundamental problem is that an equation
of motion for erosion, formulated from first principles,
is lacking.  The situation is somewhat analogous
to issues surrounding the description of the 
dynamics of granular media~\cite{jaeger96a,kadanoff99},
noting that erosion is arguably far more complex.
Nevertheless, a number of erosion equations have been proposed
ranging from deterministic~\cite{smith72,kramer92,izumi95,sinclair96}
to stochastic theories~\cite{banavar97,somfai97,pastor-satorras98,pastor98b,cieplak98a,giacometti2000u}.
Each of these models attempts to describe how eroding
surfaces evolve dynamically.
In addition, various heuristic models of both
surface and network evolution also exist.
Examples include simple lattice-based
models of erosion~\cite{kramer92,takayasu92,leheny95,caldarelli97},
an analogy to invasion percolation~\cite{stark91},
the use of optimality principles and
self-organized criticality~\cite{rodriguez-iturbe97,sun94,sun95},
and even uncorrelated 
random networks~\cite{dodds2000pa,leopold62,scheidegger67}.
Since river networks are an essential feature
of eroding landscapes, any appropriate theory of erosoion
must yield surfaces with network structures comparable
to that of the real world.
However, no model of eroding landscapes
or even simply of network evolution unambiguously reproduces
the wide range of scaling 
behavior reported for real river networks.

A considerable problem facing these
theories and models is that the values of scaling exponents
for river network scaling laws are not precisely known.
One of the issues we address in this 
work is universality~\cite{dodds2000pa,maritan96a}.
Do the scaling exponents of
all river networks belong to a unique universality
class or are there a set of classes obtained for
various geomorphological conditions?
For example, theoretical models suggest a variety
of exponent values for 
networks that are directed versus non-directed, 
created on landscapes with heterogeneous versus homogeneous erosivity
and so on~\cite{dodds2000pa,manna96,maritan96b}.
Clearly, refined measurements of scaling exponents 
are imperative if we are to be sure of any network
belonging to a particular universality class.
Moreover, given that there is no accepted theory 
derivable from simple physics, more detailed
phenomenological studies are required.

Motivated by this situation,
we perform here a detailed investigation
of the scaling properties of river networks.
We analytically characterize
fluctuations about scaling
showing that they grow with system size.
We also report significant
and ubiquitous deviations from scaling
in real river networks.
This implies surprisingly strong restrictions
on the parameter regimes where scaling holds
and cautions against
measurements of exponents that ignore such limitations.
In the case of the Mississippi basin, for example,
we find that although our study region span
four orders of magnitude in length,
scaling may be deemed valid over no more than 1.5 orders of magnitude.
Furthermore, we repeatedly find the scaling within these
bounds to be only approximate
and that no exact, single exponent can be deduced.
We show that scaling breaks down
at small scales due to the presence of linear basins
and at large scales due to
the inherent discreteness of network structure
and correlations with overall basin shape.
Significantly, this latter correlation imprints
upon river network structure
the effects and history of geology.

This paper is the first of a series of three
on river-network geometry.
Having addressed scaling laws in the present work,
we proceed in second and third articles~\cite{dodds2000ub,dodds2000uc}
to consider river network structure at a more detailed level.
In~\cite{dodds2000ub} we
examine the statistics of the ``building blocks''
of river networks, i.e.,
segments of streams and sub-networks.
In particular, we analytically connect
distributions of various kinds of stream length.
Part of this material is employed in the present
article and is a direct generalization
of Horton's laws~\cite{horton45,dodds99pa}.
In the third article~\cite{dodds2000uc}, we proceed
from the findings of~\cite{dodds2000ub}
to characterize how these building blocks fit
together.  Central to this last work is
the study of the frequency and spatial 
distributions of tributary branches along
the length of a stream and is itself a 
generalization of the descriptive picture of
Tokunaga~\cite{tokunaga66,tokunaga78,tokunaga84}.

\section{Basin allometry}
\subsection{Hack's law}

In addressing these broader issues of scaling
in branching networks, we set as
our goal to understand the river network scaling relationship
between basin area $a$ and
the length $l$ of a basin's main stream:
\begin{equation}
  \label{eq:dev.hackslaw}
  l \propto a^h.
\end{equation}
Known as Hack's law~\cite{hack57},
this relation is central to the study 
of scaling in river networks~\cite{maritan96a,dodds99pa}.
Hack's exponent $h$ is empirically found to lie in 
the range from 0.5 to 0.7~\cite{hack57,maritan96a,rigon96,gray61,mueller72,mosley73,mueller73,montgomery92,rigon98}.
Here, we postulate 
a generalized form of Hack's law
that shows good agreement with data
from real world networks.

We focus on Hack's law because
of its intrinsic interest and
also because
many interrelationships between a 
large number of scaling
laws are known and only a small subset
are understood to be independent~\cite{maritan96a,dodds99pa}.
Thus, our results for Hack's law will be in principle
extendable to other scaling laws.
With this in mind,
we will also discuss probability densities of stream
length and drainage area.

Hack's law is stated rather loosely in
equation~\req{eq:dev.hackslaw} and
implicitly involves some type of averaging
which needs to be made explicit.
It is most usually considered to be 
the relationship between \textit{mean}
main stream length and drainage area, i.e.,
\begin{equation}
  \label{eq:dev.hackslaw2}
  \avg{l} \propto a^h.
\end{equation}
Here, $\tavg{\cdot}$ denotes ensemble average
and $\tavg{l}=\tavg{l(a)}$ is the mean main stream length
of all basins of area $a$.
Typically, one performs regression analysis on
$\log{\tavg{l}}$ against $\log{a}$ to
obtain the exponent $h$. 

\subsection{Fluctuations and deviations}

\begin{figure}[tb!]
  \begin{center}
    \ifthenelse{\boolean{@twocolumn}}
    {
      \begin{tabular}{c}
        \textbf{(a)} \\
        \epsfig{file=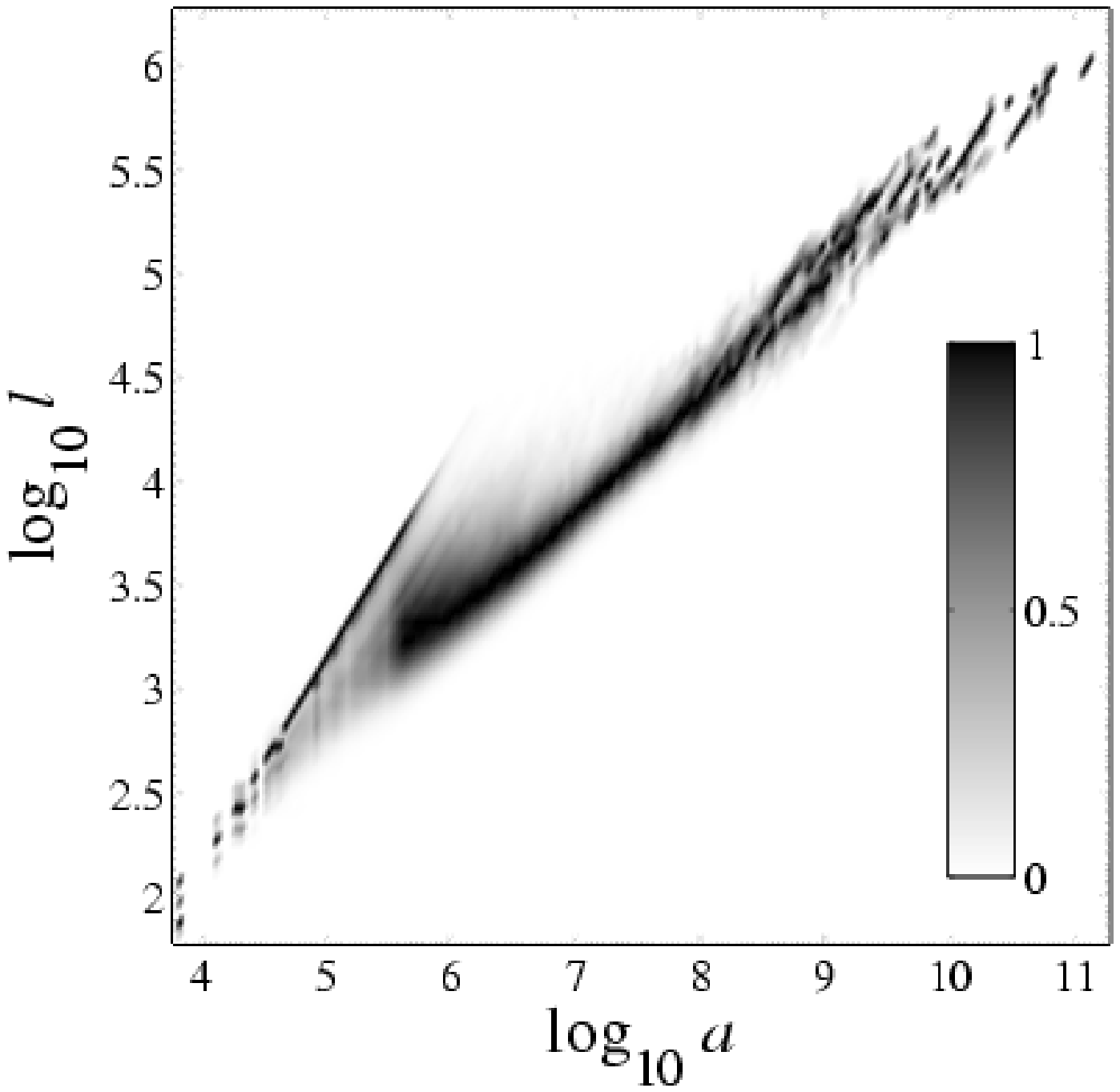,width=0.48\textwidth} \\
        \textbf{(b)} \\
        \epsfig{file=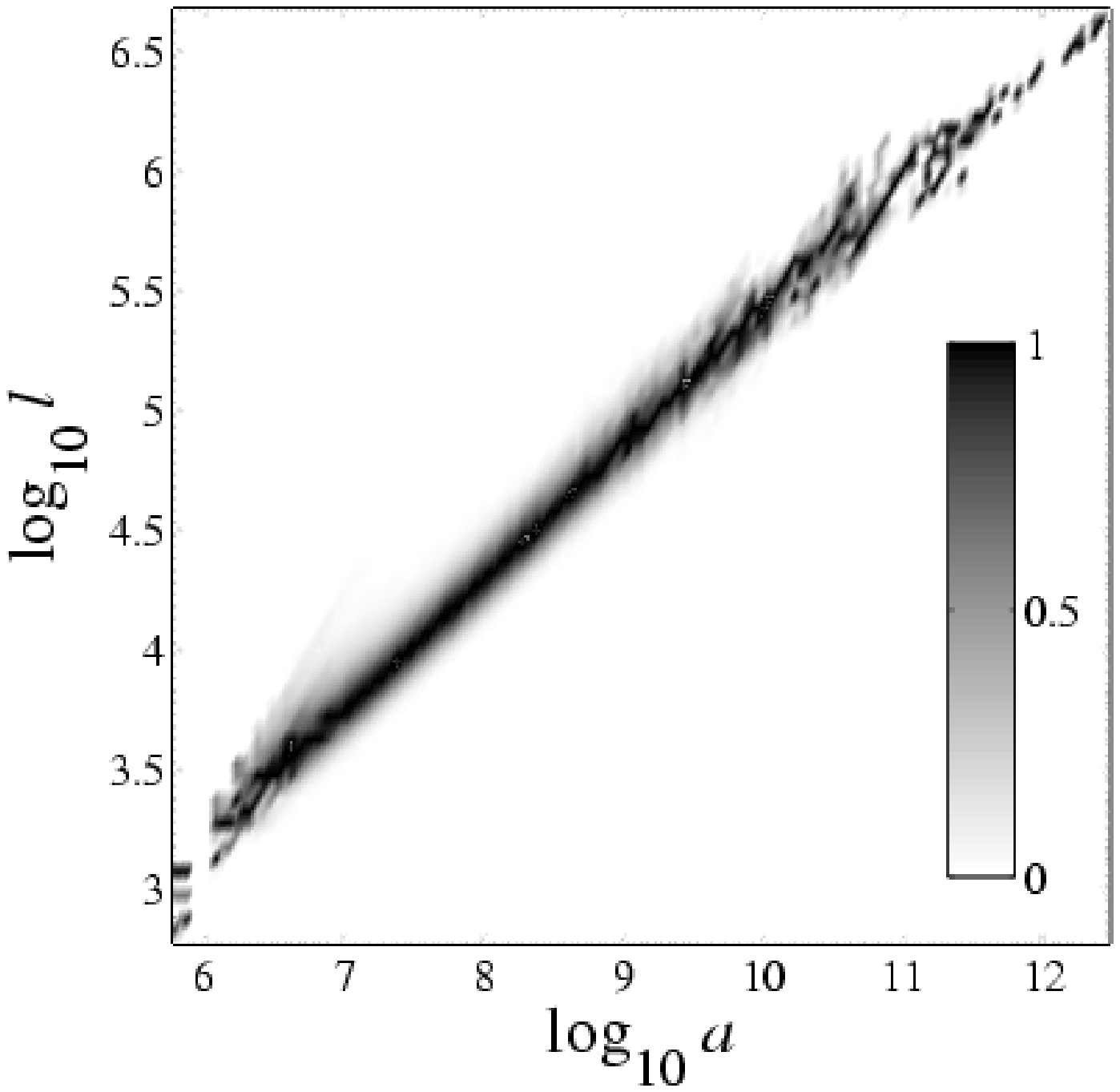,width=0.48\textwidth}
      \end{tabular}
      }
    {
      \begin{tabular}{cc}
        \textbf{(a)} & \textbf{(b)} \\
        \epsfig{file=fighack3dpcmax_gsm_kansas_noname.ps,width=0.48\textwidth} &
        \epsfig{file=fighack3dpcmax_gsm_mispi_noname.ps,width=0.48\textwidth}
      \end{tabular}
      }
    \caption[Full Hack distributions for the Kansas and Mississippi]{
      The Full Hack distribution for
      the Kansas, (a), and Mississippi, (b), river basins.  
      For each value of $a$,
      the distribution has been
      normalized along the $l$ direction
      by $\mbox{max}_{\:l} \: P(a,l)$\footcite{note:dev.fullhack}. \thnote{\input{note:dev.fullhack}}
      }
    \label{fig:dev.hack3dpcmax_kansas}
  \end{center}
\end{figure}

In seeking to understand Hack's law,
we are naturally led to wonder about the
underlying distribution that gives rise
to this mean relationship.
By considering fluctuations, we begin
to see Hack's law as an expression
of basin morphology.  What shapes of
basins characterized by $(a,l)$ are
possible and with what probability do they occur?

An important point here is that
Hack's law does not exactly specify basin shapes.
An additional connection to Euclidean dimensions
of the basin is required.  We may think
of a basin's longitudinal length $L_\parallel$
and its width $L_\perp$.
The main stream length $l$ is 
reported to scale with $L_\parallel$ as 
\begin{equation}
  \label{eq:dev.lLpar}
  l \propto L_\parallel^d,
\end{equation}
where typically $1.0 \lesssim d \lesssim 1.1$, 
\cite{maritan96a,tarboton90}.
Hence, we have $a \propto l^{1/h} \propto L_\parallel^{d/h}$.
All other relevant scaling laws exponents
can be related to the pair of exponents $(d,h)$
which therefore characterize the universality class
of a river network~\cite{dodds2000pa,dodds99pa}.
If $d/h=2$ we have that basins are self-similar
whereas if $d/h < 2$, we have that basins
are elongating.  So, while Hack's law gives
a sense of basin allometry, the fractal properties
of main stream lengths need also be known in
order to properly quantify the scaling of basin shape.

In addition to fluctuations, complementary insights
are provided by the observation 
and understanding of deviations from scaling.
We are thus interested in discerning
the regularities and quirks of the joint
probability distribution $P(a,l)$.
We will refer to $P(a,l)$ as the 
\textit{Hack distribution}.

Hack distributions for the 
Kansas river basin and 
the Mississippi river basin are given in 
Figures~\ref{fig:dev.hack3dpcmax_kansas}(a)
and~\ref{fig:dev.hack3dpcmax_kansas}(b).
Fluctuations about and 
deviations from scaling are immediately evident
for the Kansas and to a lesser extent for the Mississippi.
The first section of the paper will
propose and derive analytic forms for
the Hack distribution under the 
assumption of uniform scaling with no deviations.
Here, as well as in the following
two papers of this series~\cite{dodds2000ub,dodds2000uc},
we will motivate our results with
a random network model originally
due to Scheidegger~\cite{scheidegger67}.

We then expand our discussion
to consider deviations from exact scaling.
In the case of the 
Kansas river, 
a striking example of deviations from scaling
is the linear branch separated from
the body of the main distribution
shown in Figure~\ref{fig:dev.hack3dpcmax_kansas}(a).
This feature is
less prominent in the lower resolution
Mississippi data.
Note that this linear branch is not an artifact of the measurement
technique or data set used. 
This will be explained in our discussion
of deviations at small scales in
the paper's second section.

We then consider the more subtle deviations associated
with intermediate scales.
At first inspection, the scaling appears to 
be robust.
However, we find gradual drifts in ``exponents''
that prevent us from identifying a
precise value of $h$ and hence
a corresponding universality class.

Both distributions also
show breakdowns in scaling for
large areas and stream lengths
and this is addressed in
the final part of our section on deviations.
The reason for such deviations is partly due to 
the decrease
in number of samples and
hence self-averaging, as area and
stream lengths are increased.
However, we will show that
the direction of the deviations
depends on the overall basin shape.
We will quantify the
extent to which such deviations
can occur and the effect that
they have on measurements of Hack's 
exponent $h$.

Throughout the paper,
we will return to the Hack distributions
for the Kansas and the Mississippi rivers
as well as data obtained for
the Amazon, the Nile
and the Congo rivers.

\section{Fluctuations: an analytic form for the Hack distribution}
\label{sec:dev.flucts}

To provide some insight into the nature of 
the underlying Hack distribution,
we present a line of reasoning that will
build up from Hack's law to a scaling
form of $P(a,l)$.
First let us assume for the present discussion
of fluctuations that an exact
form of Hack's law holds:
\begin{equation}
  \label{eq:dev.hackslaw3}
  \avg{l} = \theta a^h
\end{equation}
where we have introduced the coefficient $\theta$ which
we discuss fully later on.
Now, since Hack's law is a power law,
it is reasonable to postulate a generalization
of the form
\begin{equation}
  \label{eq:dev.P(l|a)}
  \cprob{l}{a} = \frac{1}{a^h} 
  F_l \left( \frac{l}{a^h} \right).
\end{equation}
The prefactor $1/a^h$ provides the 
correct normalization and $F_l$
is the ``scaling function'' we hope to understand.
The above will be our notation
for all conditional probabilities.
Implicit in equation~\req{eq:dev.P(l|a)}
is the assumption that all moments and
the distribution itself also scale.
For example, the $q$th moment of $\cprob{l}{a}$ is
\begin{equation}
  \label{eq:dev.moms}
  \avg{l^{q}(a)} \propto a^{qh},
\end{equation}
which implies
\begin{equation}
  \label{eq:dev.mom2}
  \avg{l^{q}(a)} =  k^{-qh} \avg{l^{q}(ak)}.
\end{equation}
where $k \in \mathbb{R}$.
Also, for the distribution $\cprob{l}{a}$
it follows from equation~\req{eq:dev.P(l|a)}
that 
\begin{equation}
  \label{eq:dev.P(l|a)momscaling}
  k^h\cprob{lk^h}{ak} = \frac{k^h}{a^hk^h}
  F_l \left( \frac{lk^h}{a^hk^h} \right)
  =  \cprob{l}{a}.
\end{equation}

We note that previous investigations of Hack's law~\cite{maritan96a,rigon96}
consider the generalization in
equation~\req{eq:dev.P(l|a)}.
Rigon \etal~\cite{rigon98} also examine the behavior
of the moments of the distribution $\cprob{l}{a}$
for real networks.  Here, we will go further
to characterize the full distribution $P(a,l)$
as well as both $\cprob{l}{a}$ and $\cprob{a}{l}$.
Along these lines, Rigon \etal\ \cite{rigon96} suggest that
the function $F_l(x)$ is 
a ``finite-size'' scaling function analogous to those
found in statistical mechanics, i.e.: 
$F_l(x) \rightarrow 0$ as $x \rightarrow \infty$
and
$F_l(x) \rightarrow c$ as $x \rightarrow 0$.
However, as we will detail below, the 
restrictions on $F_l(x)$ can be made
stronger and we will postulate a
simple Gaussian form.  
More generally, $F_l(x)$ should be
a unimodal distribution that is non-zero
for an interval $[x_1,x_2]$ where $x_1 > 0$.
This is so because for any given fixed 
basin area $a$, there is a minimum 
and maximum $l$ beyond which no basin
exists.  This is also clear upon inspection
of Figures~\ref{fig:dev.hack3dpcmax_kansas}(a)
and~\ref{fig:dev.hack3dpcmax_kansas}(b).

We observe that neither drainage area nor main stream length
possess any obvious features so as to
be deemed the independent variable.
Hence, we can also view Hack's law as its inversion
$\tavg{a} \propto l^{1/h}$.  
Note that the
constant of proportionality is not necessarily
$\theta^{1/h}$ and is dependent on the 
nature of the full Hack distribution.
We thus have 
another scaling ansatz as per equation~\req{eq:dev.P(l|a)}
\begin{equation}
  \label{eq:dev.P(a|l)}
  \cprob{a}{l} = 1/l^{1/h} F_a (a/l^{1/h}).
\end{equation}

The conditional probabilities $\cprob{l}{a}$ and
$\cprob{a}{l}$ are related to the joint probability 
distribution as
\begin{equation}
  \label{eq:dev.Plinks}
  P(a,l) = P(a) \cprob{l}{a} = P(l)\cprob{a}{l},
\end{equation}
where $P(l)$ and $P(a)$ are the probability
densities of main stream length and area.
These distributions are in turn observed to
be power laws both in real world networks
and models~\cite{rodriguez-iturbe97,takayasu88,meakin91}:
\begin{equation}
  \label{eq:dev.PaPl}
  P(a) \sim N_a a^{-\tau} 
  \qquad
  \mbox{and}
  \qquad
  P(l) \sim N_l l^{-\gamma}.
\end{equation}
where $N_a$ and $N_l$ are appropriate prefactors
and the tilde indicates asymptotic agreement between both sides
for large values of the argument.
Furthermore, the exponents $\tau$ and $\gamma$ are related to 
Hack's exponent $h$ via the scaling relations~\cite{maritan96a,dodds99pa}
\begin{equation}
  \label{eq:dev.taugammah}
  \tau = 2 - h
  \qquad
  \mbox{and}
  \qquad
  \gamma = 1/h.
\end{equation}

Equations~\req{eq:dev.P(l|a)}, \req{eq:dev.P(a|l)}, 
\req{eq:dev.Plinks}, \req{eq:dev.PaPl}, and~\req{eq:dev.taugammah}
combine to give us two forms for $P(a,l)$,
\begin{equation}
  \label{eq:Palgeneral}
  P(a,l) = 1/a^{2} F(l/a^h) = 1/l^{2/h} G(a/l^{1/h}),
\end{equation}
where $x^{-2} F(x^{-h}) = G(x)$ and, equivalently,
$F(y) = y^{-2} G(y^{-1/h})$. 

\section{Random directed networks}
We will use results from the
Scheidegger model~\cite{scheidegger67}
to motivate the forms of these distributions.  
In doing so, we will also connect with some problems
in the theory of random walks.

\begin{figure}[htb!]
  \begin{center}
    \epsfig{file=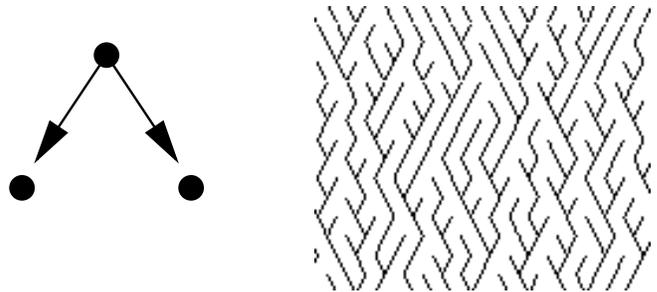,width=0.48\textwidth}
    \caption[Depiction of Scheidegger's model]{
      Scheidegger's model of random, directed networks.
      Flow is down the page and at each site, stream flow
      is randomly chosen to be in one of the two downward
      diagonals.  Stream paths and basin boundaries are
      thus discrete random walks.
      }
    \label{fig:dev.scheidmodel}
  \end{center}
\end{figure}

Scheidegger's model of river networks is
defined on a triangular lattice as
indicated by Figure~\ref{fig:dev.scheidmodel}.
Flow in the figure is directed down the page.
At each site, the stream flow direction is randomly
chosen between the two diagonal directions shown.
Periodic boundary conditions are applied
in all of our simulations.
Each site locally drains an area of $\alpha^2$,
where the lattice unit $\alpha$ is the distance between neighboring
sites, and each segment of stream has a length $\alpha$.
For simplicity, we will take $\alpha$ to be unity.
We note that connections exist between the
Scheidegger model and models of particle 
aggregation~\cite{takayasu88,huber91},
Abelian sandpiles~\cite{dhar90,dhar92,dhar99}
and limiting cases of force chain models in 
granular media~\cite{coppersmith96}.

Since Scheidegger's model is based
on random flow directions, the Hack distributions
have simple interpretations.  
The boundaries of drainage basins in the model
are random walks.  
Understanding Hack's law therefore amounts
to understanding the first collision time 
of two random walks that share the
same origin in one dimension.
If we subtract the graph of 
one walk from the other, we see that 
the latter problem is itself equivalent
to the first return problem
of random walks~\cite{feller68I}.

Many facets of the first return problem are well understood.
In particular, the probability of $n$,
the number of steps taken by a random walk until
it first returns to the origin, is asymptotically given by
\begin{equation}
  \label{eq:dev.firstreturn}
  P(n) \sim \frac{1}{\sqrt{2}\pi} n^{-3/2}.
\end{equation}
But this number of steps is also the
length of the basin $l$.  
Therefore, we have
\begin{equation}
  \label{eq:dev.Pl_sche}
  P(l) \sim \frac{2}{\pi} l^{-3/2},
\end{equation}
where because we are considering the difference
of two walks, we use $P(l) = P(n/2)|_{n=l}$.
Also, we have found the prefactor $N_l=2/\pi$.

We thus have that $\gamma=3/2$ for the
Scheidegger model.  The scaling relations
of equation~\req{eq:dev.taugammah} then
give $h=2/3$ and $\tau=4/3$.  The value
of $h$ is also readily obtained by
noting that the typical area of a basin
of length $l$ is $a \propto l\cdot l^{1/2} = l^{3/2} = l^{1/h}$
since the boundaries are random walks.

\begin{figure}[tb!]
  \begin{center}
    \epsfig{file=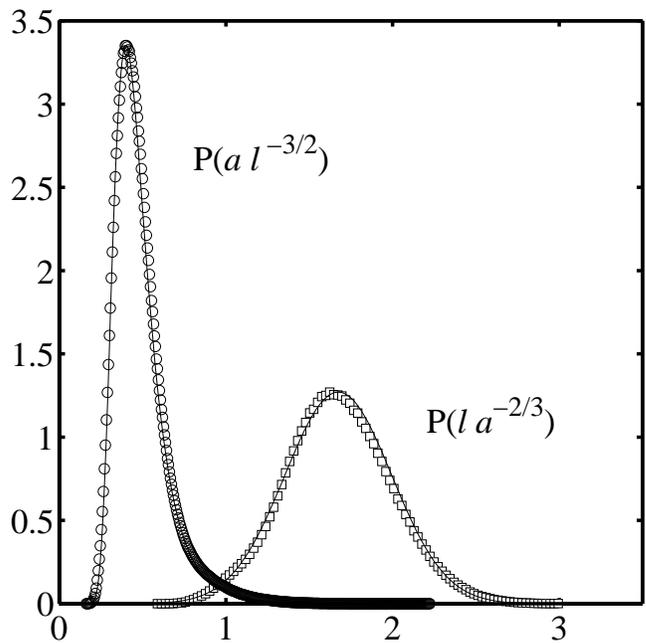,width=0.48\textwidth}
    \caption[Hack scaling functions for the Scheidegger model]{
      Cross-Sectional scaling functions of the Hack distribution for the 
      Scheidegger model with lattice constant equal to unity.
      Both distributions are normalized.
      The right distribution is for $l$ fixed
      and $a$ varying and is postulated to be a normal distribution.
      The left distribution for $a$ fixed and $l$ varying
      and is a form of an inverse Gaussian~\cite{feller68II}.  
      The data used was obtained for all sites with 
      with $l \ge 100$ and $a \ge 500$ respectively.
      Each distribution
      was obtained from ten realizations of the Scheidegger
      model on a $10^4\!\times\!10^4$ lattice.
      }
    \label{fig:dev.sche_hackscalingfn}
  \end{center}
\end{figure}

\section{Area-length distribution for random, directed networks}

Something that is less well studied
is the joint distribution of the
area enclosed by a random walk and the number
of steps to its first return.
In terms of the Scheidegger model,
this is precisely the Hack distribution.

We motivate some general results
based on observations of the Scheidegger model.
Figure~\ref{fig:dev.sche_hackscalingfn}
shows the normalized distributions $P(a l^{-3/2})$ and
$P(l a^{-2/3})$ as derived from simulations of the model.
Given the scaling ansatzes for $\cprob{l}{a}$ and $\cprob{a}{l}$
in equations~\req{eq:dev.P(l|a)} and~\req{eq:dev.P(a|l)},
we see that $P(y\!=\!l a^{-2/3}) = F_l(y)$
and $P(x\!=\!a l^{-3/2}) = F_a(x)$.

Note that we have already used Hack's law 
for the Scheidegger model with $h=2/3$
to obtain these distributions.
The results are for ten realizations of
the model on a $10^4$ by $10^4$ lattice,
taking $10^7$ samples from each of the ten instances.
For $\cprob{a}{l}$, only sites
where $l \ge 100$ were taken, and similarly,
for $\cprob{l}{a}$, only sites
where $a \ge 500$ were included in the histogram.

We postulate that the distribution $P(y=l a^{-2/3})$
is a Gaussian having the form
\begin{equation}
  \label{eq:dev.P(l|a)_sche}
  \cprob{l}{a} = \frac{1}{\sqrt{2\pi} a^{2/3} \eta}
  \exp\{-(l a^{-2/3}-\theta)^2/2\eta^2\}
\end{equation}
We estimate the mean of $F_l$ to be $\theta \simeq 1.675$
(this is the same $\theta$ as found in equation~\req{eq:dev.hackslaw3})
and the standard deviation to be $\eta \simeq 0.321$.
The fit is shown in Figure~\ref{fig:dev.sche_hackscalingfn}
as a solid line.
The above equation agrees with the form of the scaling ansatz
of equation~\req{eq:dev.P(l|a)} and we now have the assertion that
$F_l$ is a Gaussian defined by the 
two parameters $\theta$ and $\eta$.

Note that the $\theta$ and
$\eta$ are coefficients for the actual mean and standard deviation.
In other words, for fixed $a$, the mean of $\cprob{l}{a}$ is 
$\theta a^{2/3}$ and its standard deviation
is $\eta a^{2/3}$.  Having observed their context,
we will refer to $\theta$ and $\eta$
as the Hack mean coefficient and Hack standard deviation coefficient.

From this starting point we can
create $P(a,l)$ and $\cprob{a}{l}$, the latter providing
a useful test.  Since $P(a) \sim N_a a^{-\tau} = N_a a^{-4/3}$,
as per equation~\req{eq:dev.PaPl},
we have
\begin{eqnarray}
  \label{eq:dev.Pal_sche}
  P(a,l) & = & \frac{N_a}{a^{4/3}}
  \frac{1}{\sqrt{2\pi} a^{2/3} \eta}
  \exp\{-(l a^{-2/3}-\theta)^2/2\eta^2\}, \nonumber \\
  & = & 
  \frac{N_a}{\sqrt{2\pi} a^{2} \eta}
  \exp\{-(l a^{-2/3}-\theta)^2/2\eta^2\}.
\end{eqnarray}
As expected, we observe
the form of equation~\req{eq:dev.Pal_sche}
to be in accordance with that of equation~\req{eq:Palgeneral}.
Note that the scaling function $F$ (and equivalently $G$) is defined
by the three parameters $\theta$, $\eta$ and $N_a$, the latter
of which may be determined in terms of the former as we
will show below.
Also, since we expect all scaling functions to be only asymptotically
correct, we cannot use equation~\req{eq:dev.Pal_sche}
to find an expression for the normalization $N_a$.
Equation~\req{eq:dev.Pal_sche} ceases to be valid
for small $a$ and $l$.
However, we will be able to do so once we have $\cprob{a}{l}$
since we are able to presume $l$ is large
and therefore that the scaling form is exact.
Using equation~\req{eq:dev.Pl_sche}
and the fact that $\cprob{a}{l} = P(a,l)/P(l)$
from equation~\req{eq:dev.Plinks}
we then have 
\begin{eqnarray}
  \label{eq:dev.sche_P(a|l)}
  \lefteqn{\cprob{a}{l}
  = \frac{\pi l^{3/2}}{2}
  \frac{N_a}{\sqrt{2\pi} a^{2} \eta}
  \exp\{-(l/a^{2/3}-\theta)^2/2\eta^2\},} \nonumber \\
  & = & \frac{N_a \sqrt{\pi} l^{3/2}}{2^{3/2}a^{2} \eta}
  \exp\{-(l/a^{2/3}-\theta)^2/2\eta^2\}, \nonumber \\
  & = & \frac{1}{l^{3/2}}
  \frac{N_a \sqrt{\pi}}{2^{3/2}\eta} (a/l^{3/2})^{-2} \nonumber\\
  & & \times \exp\{-((a/l^{3/2})^{-2/3}-\theta)^2/2\eta^2\}.
\end{eqnarray}
In rearranging the expression of $\cprob{a}{l}$,
we have made clear that its form matches that
of equation~\req{eq:dev.P(l|a)}.

A closed form expression 
for the normalization factor $N_a$ may now be determined
by employing the fact that $\int_{a=0}^\infty \dee{a} \cprob{a}{l} = 1$.
\begin{eqnarray}
  \label{eq:dev.Nacalc}
  \lefteqn{ 1 = \int_{a=0}^\infty \dee{a} \cprob{a}{l},} \nonumber \\
  & = & \int_{a=0}^\infty \frac{\dee{a}}{l^{3/2}}
  \frac{N_a \sqrt{\pi}}{2^{3/2}\eta} (a/l^{3/2})^{-2} \nonumber \\
  & & \times
  \exp\{-((a/l^{3/2})^{-2/3}-\theta)^2/2\eta^2\}, \nonumber \\
  & = & \frac{N_a \sqrt{3\pi}}{2^{5/2}\eta}
  \int_{u=0}^\infty \dee{u}
  u^{1/2} \exp\{-(u-\theta)^2/2\eta^2\},
\end{eqnarray}
where we have used the substitution $a/l^{3/2}=u^{-3/2}$
and hence also $l^{-3/2}\postdee{a}=(-3/2)u^{-5/2}\postdee{u}$.
We therefore have
\begin{equation}
  \label{eq:dev.Nacalc2}  
   N_a =  \frac{2^{5/2}\eta}{\sqrt{3\pi}}
  \left[ \int_{u=0}^\infty \dee{u}
  u^{1/2} \exp\{-(u-\theta)^2/2\eta^2\} \right]^{-1}.
\end{equation}

We may thus
write down all of the scaling functions $F_l$, $F_a$, $F$ and $G$
for the Scheidegger model:
\begin{eqnarray}
  \label{eq:dev.sche_scfnsFl}
  F_l(z) & = & \frac{1}{\sqrt{2\pi} \eta}
  \exp\{-(z-\theta)^2/2\eta^2\}, \\
  \label{eq:dev.sche_scfnsFa}
  F_a(z) & = & \frac{N_a \sqrt{\pi}}{2^{3/2}\eta} z^{-2}
  \exp\{-(z^{-2/3}-\theta)^2/2\eta^2\}, \\
  \label{eq:dev.sche_scfnsF}
  F(z) & = & \frac{N_a}{\sqrt{2\pi} \eta}
  \exp\{-(z-\theta)^2/2\eta^2\}, \ \mbox{and}\\
  \label{eq:dev.sche_scfnsG}
  G(z) & = & \frac{2N_a}{\sqrt{\pi}2^{3/2}\eta} z^{-2}
  \exp\{-(z^{-2/3}-\theta)^2/2\eta^2\}.
\end{eqnarray}

Recall that all of these forms rest on the assumption that
$F_l(z)$ is a Gaussian.
In order to check this assumption, we
return to Figure~\ref{fig:dev.sche_hackscalingfn}.
The empirical distribution $P(z=a/l^{3/2})$ is
shown on the left marked with circles.
The solid line through these
points is $F_a(z)$ as given above in 
equation~\req{eq:dev.sche_scfnsFl}.
There is an excellent match so we may be
confident about our proposed form for $F_a(z)$.
We note that the function $F_a(z)$ may be thought of as a
fractional inverse Gaussian distribution, the
inverse Gaussian being a well known distribution
arising in the study of first passage times
for random walks~\cite{feller68I}.
It is worth contemplating the peculiar form
of $\cprob{a}{l}$ in terms of first return random walks.
Here, we have been able to postulate the functional
form of the distribution of areas bound by random
walks that first return after $n$ steps.
If one could understand the origin of the
Gaussian and find analytic expressions
for $\theta$ and $\eta$, then the problem would
be fully solved.

\section{Area-length distribution extended to real networks}
We now seek to extend these results 
for Scheidegger's model to real world networks.
We will look for the same functional forms
for the Hack distributions that we have found above.
The conditional probability 
distributions pertaining to Hack's law 
take on the forms
\begin{equation}
  \cprob{l}{a}= 1/a^{-h} F_l(l a^{-h})
   =  \frac{a^{-h}}{\sqrt{2\pi} \eta}
  \exp\{-(l a^{-h}-\theta)^2/2\eta^2\},
  \label{eq:dev.scfnsFl}
\end{equation}
and 
\ifthenelse{\boolean{@twocolumn}}
{
  \begin{eqnarray}
    \lefteqn{\cprob{a}{l} = l^{-1/h} F_a(a l^{-1/h})} \nonumber \\
    & = & l^{-1/h} \frac{N_a}{\sqrt{2\pi} N_l \eta} 
    (a l^{-1/h})^{-2}
    \exp\{-((a l^{-1/h})^{-h}-\theta)^2/2\eta^2\}, \nonumber \\
    \label{eq:dev.scfnsFa}
  \end{eqnarray}
}
{
  \begin{equation}
    \cprob{a}{l} = l^{-1/h} F_a(a l^{-1/h})
    = l^{-1/h} \frac{N_a}{\sqrt{2\pi} N_l \eta} 
    (a l^{-1/h})^{-2}
    \exp\{-((a l^{-1/h})^{-h}-\theta)^2/2\eta^2\},
    \label{eq:dev.scfnsFa}
  \end{equation}
}
and the full Hack distribution is given by
\begin{eqnarray}
  P(a,l) & = & a^{2} F(l a^{-h}) = l^{-2/h} G(a l^{-1/h}) \nonumber \\
   & = & \frac{N_a}{\sqrt{2\pi} \eta a^2}
  \exp\{-(l a^{-h}-\theta)^2/2\eta^2\}.
  \label{eq:dev.scfnsF}
\end{eqnarray}
with $N_a$ determined by equation~\req{eq:dev.Nacalc2}.
The three parameters 
$h$, the Hack exponent, $\theta$, the Hack mean coefficient, and
$\eta$, the Hack standard deviation coefficient,
are in principle
landscape dependent.  
Furthermore, in $\eta$ we
have a basic measure of fluctuations 
in the morphology of basins.

\begin{figure}[tb!]
  \begin{center}
    \epsfig{file=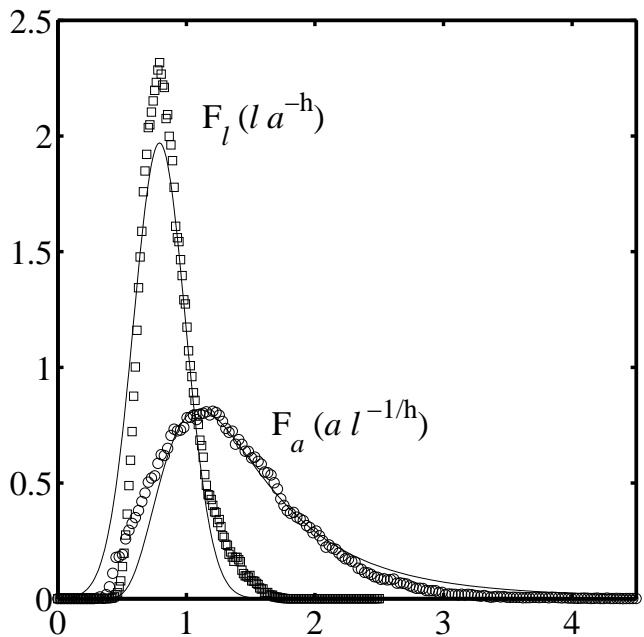,width=0.48\textwidth}
    \caption[Hack Scaling functions for the Mississippi]{
      Cross-sectional scaling functions of the Hack 
      distribution for the 
      Mississippi.
      The estimates used for Hack's exponent
      are $h = 0.55$ and $h=0.50$ respectively, 
      the determination of which is
      discussed in section~\ref{sec:dev.deviations}.
      The fits indicated by the smooth curves to the data are made 
      as per the Scheidegger model in Figure~\ref{fig:dev.sche_hackscalingfn}
      and according to 
      equations~\req{eq:dev.scfnsFl} and~\req{eq:dev.scfnsFl}.
      The values of the Hack mean coefficient and
      standard deviation coefficient are
      estimated to be $\theta \simeq 0.80$
      and $\eta \simeq 0.20$.      
      }
    \label{fig:dev.hack_scfns_mispi}
  \end{center}
\end{figure}

\begin{figure}[tb!]
  \begin{center}
    \epsfig{file=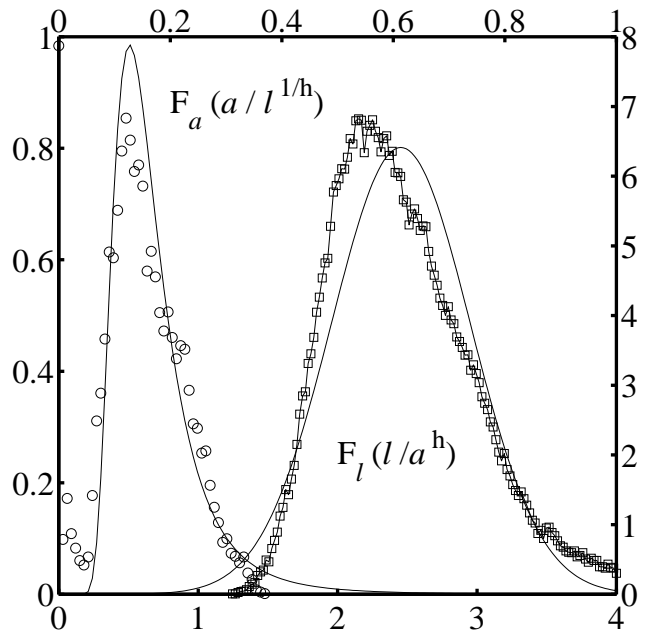,width=0.48\textwidth} 
    \caption[Hack Scaling functions for the Nile]{
      Cross-sectional scaling functions 
      of the Hack distribution for the 
      Nile.  The top and right axes correspond
      to $F_a$ and the bottom and left to $F_l$.
      The Hack exponent used is $h=0.50$
      and the values of the Hack mean coefficient and
      standard deviation coefficient are estimate to be
      $\theta \simeq 2.45$ and 
      $\eta \simeq 0.50$\footcite{note:dev.nilecongo}. \thnote{\input{note:dev.nilecongo}}
      }
    \label{fig:dev.hack_scfns_nile}
  \end{center}
\end{figure}

Figures~\ref{fig:dev.hack_scfns_mispi}
and~\ref{fig:dev.hack_scfns_nile}
present Hack scaling functions for the Mississippi
and Nile river basins.
These Figures is to be compared with the 
results for the Scheidegger model
in Figure~\ref{fig:dev.sche_hackscalingfn}.

For both rivers, Hack's exponent
$h$ was determined first from a
stream ordering analysis 
(we discuss stream ordering later 
in Section~\ref{subsec:dev.large}).
Estimates of the parameters $\theta$
and $\eta$ were then made using the
scaling function $F_l$  
presuming a Gaussian form.

We observe the Gaussian fit for the 
Mississippi is more satisfactory
than that for the Nile.
These fits are not rigorously made because
even though we have chosen data ranges
where deviations (which we address in
the following section) are minimal,
deviations from scaling do still skew
the distributions.
The specific ranges used
to obtain $F_l$ and $F_a$ 
respectively are for the Mississippi:
$8.5  < \log_{10}a <9.5 $ and $4.75  < \log_{10}l <6 $,
and for the Nile:
$9  < \log_{10}a <11 $ and $5.5  < \log_{10}l <6 $
(areas are in km$^2$ and lengths km).
Furthermore, we observe that the estimate of $h$ has an
effect in the resulting forms
of $F_l$ and $F_a$.  
Nevertheless, here we are attempting to
capture the essence of the generalized
form of Hack's law in real networks.

We then use the parameters $h$, $\theta$
and $\eta$ and equation~\req{eq:dev.scfnsFa}
to construct our theoretical $F_a$,
the smooth curves in Figures~\ref{fig:dev.hack_scfns_mispi} 
and~\ref{fig:dev.hack_scfns_nile}.
As for the Scheidegger model data
in Figure~\ref{fig:dev.sche_hackscalingfn},
we see in both examples
approximate agreement between the measured
$F_a$ and the one 
predicted from the form of $F_l$.
Table~\ref{tab:dev.hackdistparams}
shows estimates of $h$, $\theta$ and $\eta$
for the five major river basins studied.

\begin{table}[htb!]
  \begin{center}
    \begin{tabular}{lccc}
      River network & $\theta$ & $\eta$ & $h$ \\
      Mississippi & 0.80 & 0.20 & 0.55 \\
      Amazon & 1.90 & 0.35 & 0.52 \\
      Nile & 2.45 & 0.50 & 0.50 \\
      Kansas & 0.70 & 0.15 & 0.57 \\
      Congo & 0.89 & 0.18 & 0.54
    \end{tabular}
    \caption[Estimates of Hack distribution parameters for real river networks]{
      Estimates of Hack distribution parameters for real river networks.
      The scaling exponent $h$ is Hack's exponent.
      The parameters $\theta$ and $\eta$ are coefficients of the mean
      and standard deviation of the conditional
      probability density function $\cprob{l}{a}$ and 
      are fully discussed in the text\footcite{note:dev.amazon}. \thnote{\input{note:dev.amazon}}
      }
    \label{tab:dev.hackdistparams}
  \end{center}
\end{table}

Given our reservations about the precision of
these values of $\theta$ and $\eta$, we are nevertheless
able to make qualitative distinctions.  
Recalling that $\tavg{l} = \theta a^h$, we see
that, for fixed $h$,
higher values of $\theta$ indicate relatively
longer stream lengths for a given area and 
hence longer and thinner basins.  
The results therefore suggest 
the Nile, and to a lesser
degree the Amazon, have basins with thinner profiles
than the Congo and, in particular, the Mississippi and Kansas.
This seems not unreasonable since the Nile is
a strongly directed network constrained within
a relatively narrow overall shape.
This is somewhat in spite of the fact
that the shape of an overall river basin
is not necessarily related to its internal
basin morphology, an observation we 
will address later in Section~\ref{subsec:dev.large}.

The importance of $\theta$ is tempered
by the value of $h$.  
Hack's exponent affects not only the absolute
measure of stream length for a given area but
also how basin shapes change with increasing area.
So, in the case of the Kansas the higher
value of $h$ suggests basin profiles thin
with increasing size.  This is in keeping
with overall directedness of the network.
Note that our measurements of the fractal
dimension $d$ of stream lengths for the Kansas
place it to be $d = 1.04 \pm 0.02$.
Therefore, $d/h \simeq 1.9 < 2$ and
elongation is still expected when
we factor in the scaling of $l$ with $L_\parallel$.

The Nile and Amazon also 
all have relatively high $\eta$
indicating greater fluctuations in basin shape.
In comparison, the Mississippi, Kansas and Congo
appear to have less variation.  
Note that the variability of the Kansas is 
in reasonable agreement with 
that of the whole Mississippi river network
for which it is a sub-basin.

Finally, regardless of the actual form of 
the distribution underlying Hack's law, fluctuations
are always present and an estimate of their extent is
an important measurement.  
Thus, the Hack mean and
standard deviation coefficients, $\theta$ and $\eta$,
are suggested to be of sufficient worth so as to be
included with any measurement of the Hack exponent $h$.

\section{Deviations from scaling}
\label{sec:dev.deviations}

In generalizing Hack's law,
we have sought out regions of robust
scaling, discarding ranges where
deviations become prominent.
We now bring our attention to
the nature of the deviations themselves.

We observe three 
major classes of deviations
which we will define by the scales
at which they occur:
small, intermediate and large.
Throughout the following sections 
we primarily consider deviations from 
the mean version of Hack's law, $\tavg{l} = \theta a^h$,
given in equation~\req{eq:dev.hackslaw3}.
Much of the understanding we gain from this will
be extendable to deviations for higher moments.

\begin{figure}[tp!]
  \begin{center}
    \ifthenelse{\boolean{@twocolumn}}
    {
      \begin{tabular}{c}
        \textbf{(a)} \\
        \epsfig{file=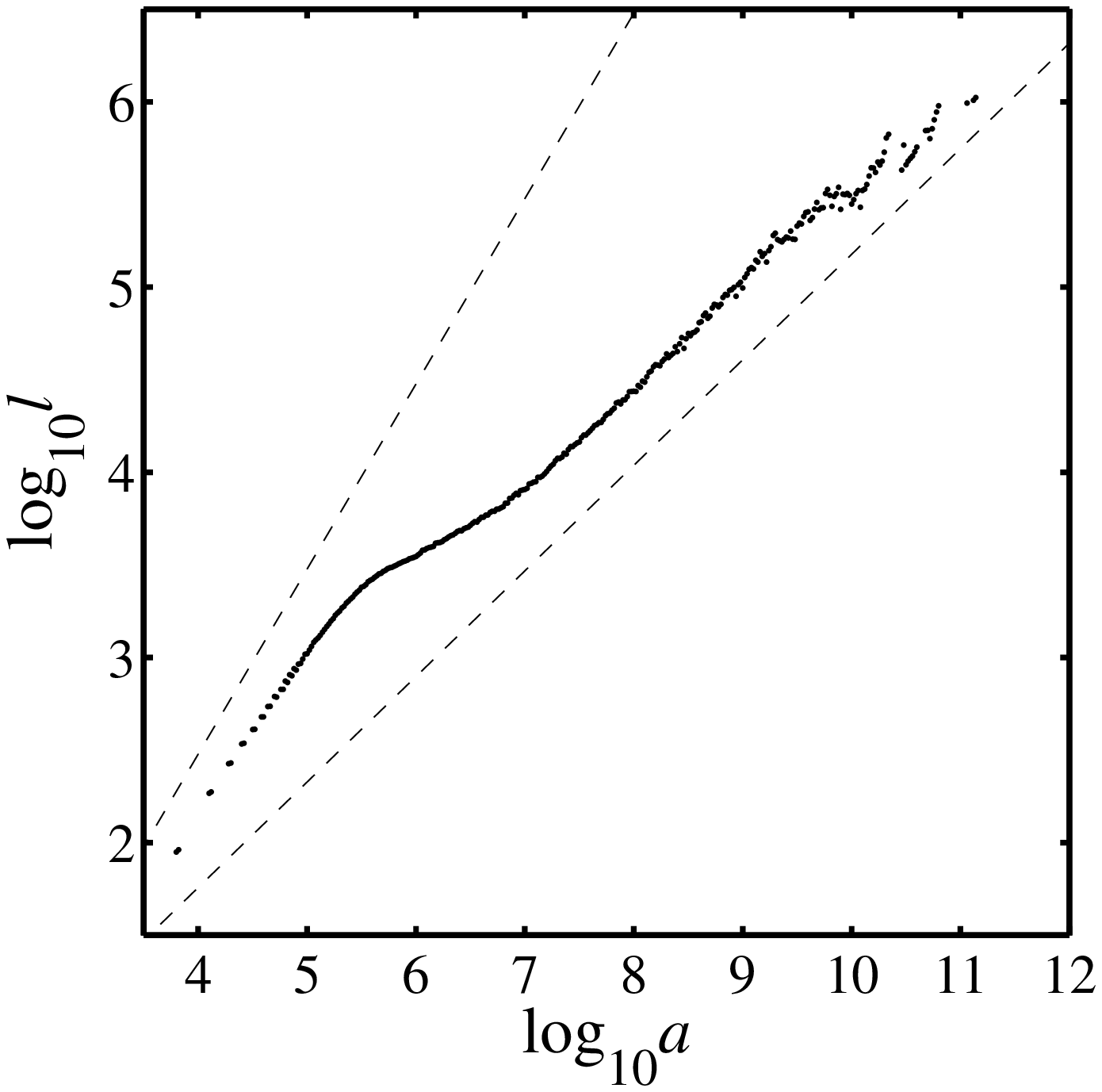,width=0.48\textwidth} \\
        \textbf{(b)} \\
        \epsfig{file=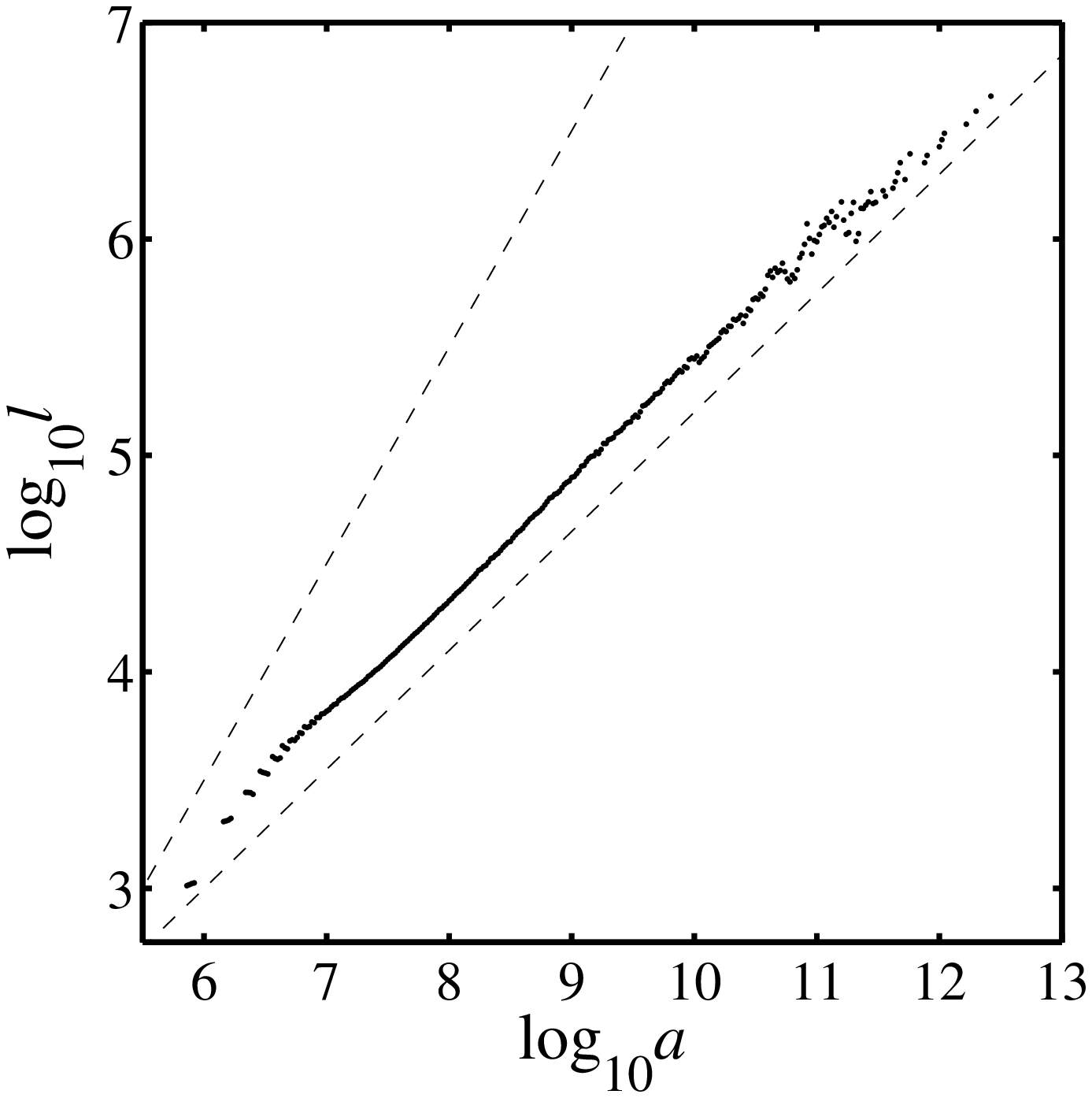,width=0.48\textwidth}
      \end{tabular}
      }
    {
      \begin{tabular}{cc}
        \textbf{(a)} & \textbf{(b)} \\
        \epsfig{file=figmeanhackdist_kansas_noname.ps,width=0.48\textwidth} &
        \epsfig{file=figmeanhackdist_mispi_noname.ps,width=0.48\textwidth}
      \end{tabular}
      }
    \caption[Mean Hack's law for the Kansas and Mississippi]{
      The mean version of Hack's law for the Kansas, (a), and
      the Mississippi, (b).
      The units of lengths and areas are meters and square meters.
      These are calculated from the full Hack distributions
      shown in Figures~\ref{fig:dev.hack3dpcmax_kansas}(a)
      and~\ref{fig:dev.hack3dpcmax_kansas}(b)
      by finding $\tavg{l}$ for each value of basin area $a$.
      Area samples are taken 
      every 0.02 orders of magnitude in logarithmic space.
      The upper dashed lines represent a slope of unity in both plots
      and the lower lines the Hack exponents
      $0.57$ and $0.55$ for the Kansas and Mississippi respectively.
      For the Kansas, there is a clear deviation for small area which
      rolls over into a region of very slowly changing
      derivative before breaking up at large scales.
      Deviations from scaling are present for the lower
      resolution dataset of the Mississippi
      but to a lesser extent.
      }
    \label{fig:dev.meanhackdist_kansas}
  \end{center}
\end{figure}

To provide an overview
of what follows, examples of mean-Hack distributions
for the Kansas river and the Mississippi are
shown in Figures~\ref{fig:dev.meanhackdist_kansas}(a)
and~\ref{fig:dev.meanhackdist_kansas}(b).
Hack's law for the Kansas river exhibits a marked
deviation for small areas, starting
with a near linear relationship between
stream length and area.  
A long crossover region of several orders of magnitude
in area then 
leads to an intermediate scaling regime
wherein we attempt to determine the Hack exponent $h$.
In doing so, we show that
such regions of robust scaling are 
surprisingly limited for river network quantities.
Moreover, we observe
that, where present, scaling is only approximate
and that no exact exponents can be ascribed to
the networks we study here.
It follows that the identification 
of universality 
classes based on empirical evidence
is a hazardous step.

Finally, the approximate scaling of this intermediate
region then gives way to a break down in scaling at larger scales
due to low sampling and correlations with basin shape.
The same deviations are present in the 
relatively coarse-grained Mississippi data but are less pronounced.

\section{Deviations at small scales}
\label{subsec:dev.small}

\begin{figure}[tb!]
  \begin{center}
    \epsfig{file=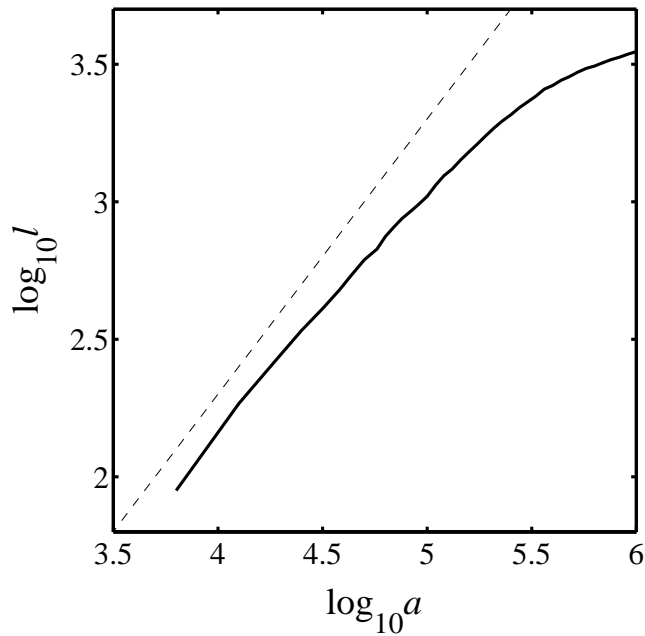,width=0.48\textwidth}
    \caption[Small scale linearity of Hack's law for the Kansas]{
      The linearity of Hack's law at small scales for the
      Kansas river.  
      The linear regime enters a crossover region
      after almost 1.5 orders of magnitude in area.
      This is an expanded detail of the mean Hack's law 
      given in Figure~\ref{fig:dev.meanhackdist_kansas}(a)
      and areas and lengths are in square meters and meters.
      }
    \label{fig:dev.meanhackdist2_kansas}
  \end{center}
\end{figure}

At small scales,
we find the mean-Hack
distribution to follow a
linear relationship, i.e., $l \propto a^1$.
This feature is most evident
for the Kansas river
as shown in Figure~\ref{fig:dev.meanhackdist_kansas}(a)
and in more detail in Figure~\ref{fig:dev.meanhackdist2_kansas}.
The linear regime persists for nearly 1.5 orders of
magnitude in basin area.
To a lesser extent,
the same trend is apparent
in the Mississippi data, 
Figure~\ref{fig:dev.meanhackdist_kansas}(b).

Returning to the full Hack distribution
of Figures~\ref{fig:dev.hack3dpcmax_kansas}(a)
and~\ref{fig:dev.hack3dpcmax_kansas}(b),
we begin to see the origin of this linear regime.
In both instances, a linear branch
separates from the body
of the main distribution.
Since $l$ cannot grow faster than $a^1$, 
the linear branch marks an upper bound 
on the extent of the distribution in $(a,l)$ coordinates.
When averaged to
give the mean-Hack distribution, this linear
data dominates the result for small scales.

\begin{figure}[tb!]
  \begin{center}
    \ifthenelse{\boolean{@twocolumn}}
    {
      \epsfig{file=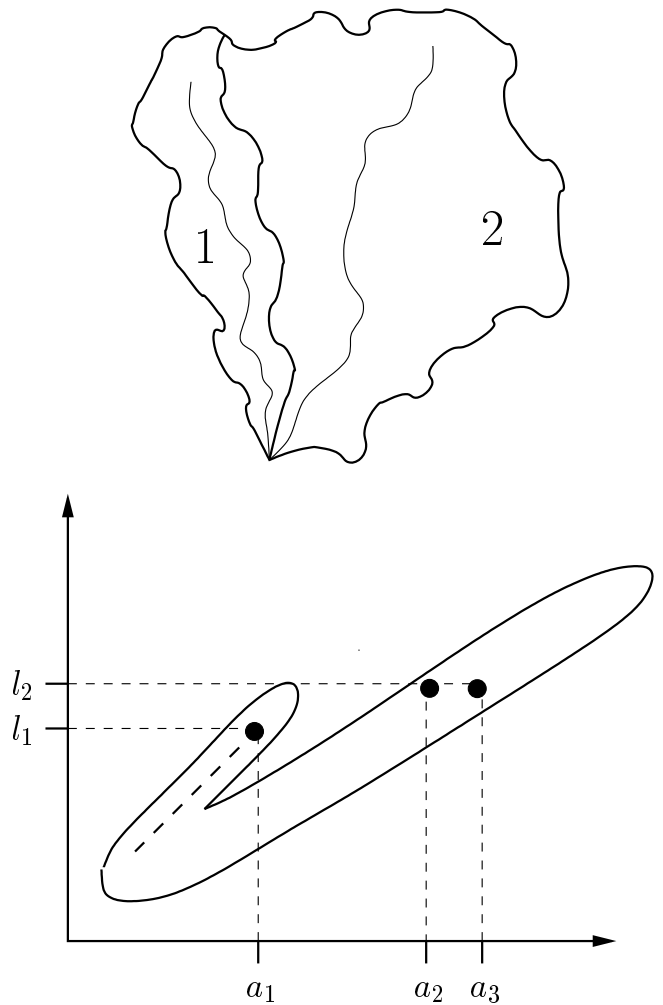,width=.48\textwidth}
      }
    {
      \epsfig{file=hackjump.ps,width=\textwidth}
      }
    \caption[Origin of the linear branch in the Hack distribution]{
      Origin of the linear branch in the Hack distribution.
      The sub-basins labeled $1$ and $2$ depicted on the left have
      areas $a_1$ and $a_2$ and lengths $l_1$ and $l_2$.
      These sub-basins combine to form a basin of area $a_3=a_1+a_2$ and 
      main stream length $l_3=\max{l_1,l_2}=l_2$.
      Since sub-basin $1$ is a linear sub-network (a valley)
      the pair $(a_1,l_1)$ lie along the linear branch of the 
      Hack distribution as shown on the right.  Points along
      the main stream of sub-basin $1$ lie along the dashed
      line leading to the coordinate $(a_1,l_1)$.
      On combining with the second basin, the jump in the
      resultant area creates a jump from $(a_1,l_1$
      to $(a_3,l_2)$ in the main body of the
      Hack distribution.
      }
    \label{fig:dev.hackjump}
  \end{center}
\end{figure}

We find this branch evident in all Hack distributions.
It is not an artifact of resolution and in fact
becomes more pronounced with increased map precision.
The origin of this linear branch is simple: data points 
along the branch correspond to positions in narrow
sub-networks, i.e., long, thin ``valleys.''
To understand the separation of this linear
branch from the main body of the distribution,
consider Figure~\ref{fig:dev.hackjump}
which depicts a stream draining such a valley
with length and area $(l_1,a_1)$ that meets a stream
from a basin with characteristics $(l_2,a_2)$.  
The area and length of the basin formed at
this junction is thus $(a_3,l_3)=(a_1+a_2,\max(l_1,l_2))$
(in the Figure, $\max(l_1,l_2)=l_2$).
The greater jump in area moves 
the point across into the main
body of the Hack distribution, 
creating the separation of the linear branch.

In fact, the full Hack distribution is itself
comprised of many such linear segments.  
As in the above example,
until a stream does not meet any streams of 
comparable size, then its area and length
will roughly increase in linear fashion.
When it does meet such a stream, there
is a jump in area and 
the trace of a new linear segment is 
started in the distribution.
We will see this most clearly later on when
we study deviations at large scales.

For very fine scale maps, on the order of meters,
we might expect to pick up the scale of
the unchannelized, convex regions
of a landscape, i.e., ``hillslopes''~\cite{dietrich98}.
This length scale represents the 
typical separation of branches at a
network's finest scale.
The computation of stream networks for
these hillslope regions would result in largely
non-convergent (divergent or parallel) flow.
Therefore, we would have linear ``basins''
that would in theory contribute to the linear branch
we observe~\cite{dodds2000pa}.  
Potentially, the crossover in Hack's
law could be used as a determinant of hillslope
scale, a crucial parameter 
in geomorphology~\cite{dietrich98,dietrich93a}.
However, when long, thin network structures
are present in a network, this hillslope scale
is masked by their contribution.

Whether because of the hillslope scale
or linear network structure, we see that
at small scales, Hack's law will show a crossover in
scaling from $h=1$ to a lower exponent.
The crossover's position depends on the extent
of linear basins in the network.
For example, in the Kansas River basin, the crossover occurs when
$(l,a)=(4\times10^3 \mbox{m},10^7 \mbox{m}^2)$.
Since increased map resolution can only 
increase measures of length,
the crossover's position must occur
at least at such a length scale
which may be many orders of magnitude
greater than the scale of the map.

However, the measurement
of the area of such linear basins will potentially grow
with coarse-graining.
Note that for the Mississippi mean-Hack
distribution, the crossover begins around
$a=10^{6.5}$ m$^2$ whereas for the Kansas,
the crossover initiates near $a=10^{5.5}$ m$^2$
but the ends of the crossovers in both cases
appear to agree, occurring at around $a=10^{7}$ m$^2$.
Continued coarse-graining will
of course eventually destroy all statistics
and introduce spurious deviations.
Nevertheless, we see here that the 
deviation which would only be 
suggested in the Mississippi
data is well confirmed in the finer-grain
Kansas data.

\section{Deviations at intermediate scales}
\label{subsec:dev.intermediate}

As basin area increases,
we move out of the linear regime, observing
a crossover to what would be 
considered the normal scaling region of Hack's law.
We detail our attempts to measure the Hack exponents
for the Kansas and Mississippi examples.
Rather than relying solely on a single regression
on a mean-Hack distribution, we
employ a more precise technique
that examines the distribution's derivative.
As we will show, we will not be able to
find a definite value for the Hack exponent
in either case,
an important result
in our efforts to determine whether or
not river networks belong to specific
universality classes.  

\begin{figure}[tp!]
  \begin{center}
    \ifthenelse{\boolean{@twocolumn}}
    {
    \begin{tabular}{c}
      \textbf{(a)} \\
      \epsfig{file=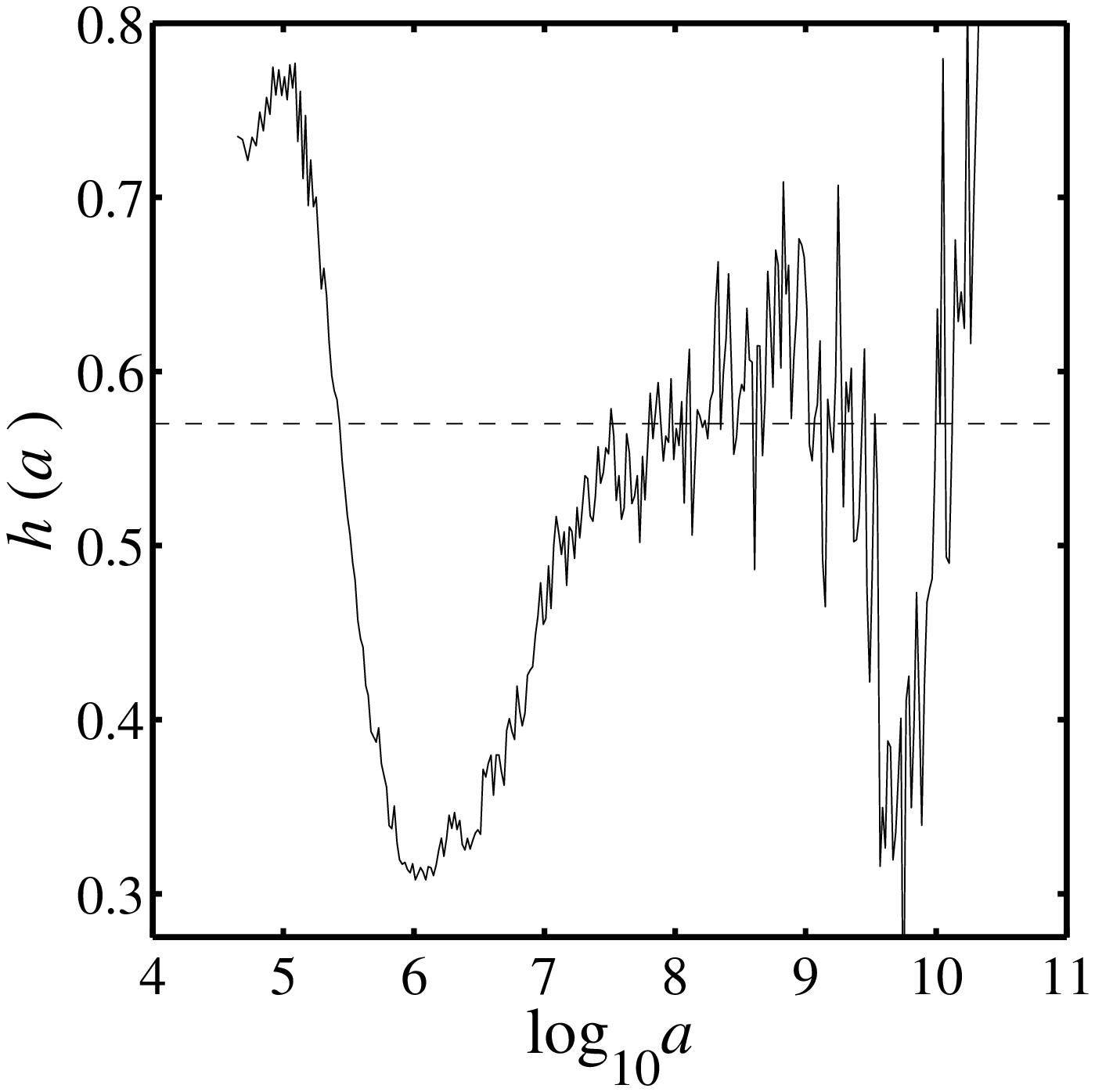,width=0.48\textwidth} \\
      \textbf{(b)} \\
      \epsfig{file=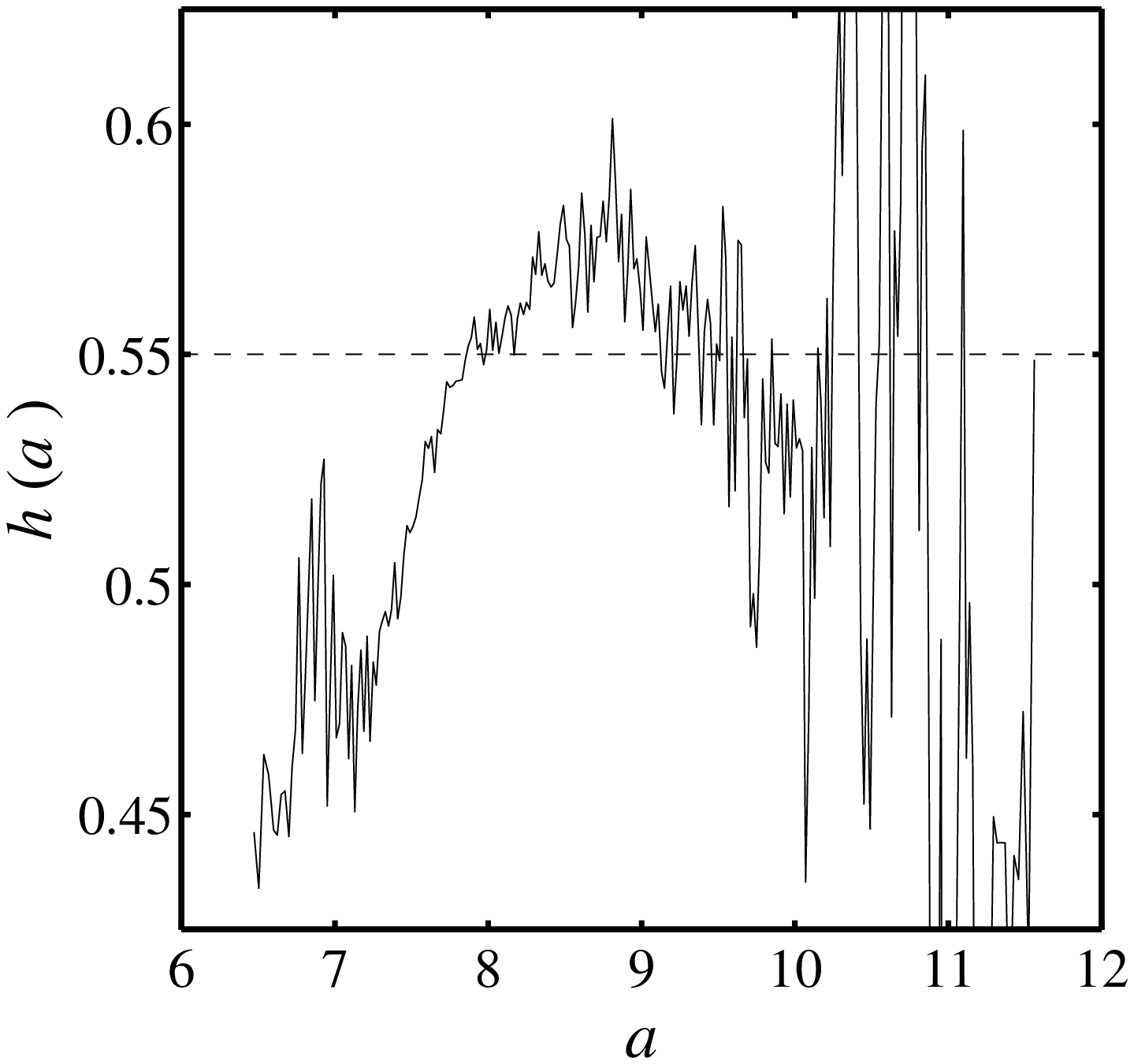,width=0.48\textwidth}
    \end{tabular}
    }
    {
    \begin{tabular}{cc}
      \textbf{(a)} & \textbf{(b)} \\
      \epsfig{file=figdiffmeanhack_kansas_noname.ps,width=0.48\textwidth}
      &
      \epsfig{file=figdiffmeanhack_mispi_noname.ps,width=0.48\textwidth}
    \end{tabular}
    }
    \caption[Variation in Hack's exponent for the Kansas and Mississippi]{
      Variation in Hack's exponent for the Kansas, (a),
      and the Mississippi, (b).  The area $a$ is in m$^2$.
      The plots are derivatives of the mean-Hack distributions given in
      Figures~\ref{fig:dev.meanhackdist_kansas}(a)
      and~\ref{fig:dev.meanhackdist_kansas}(b).
      Both derivatives have been smoothed by taking running averages
      over $0.64$ orders of magnitude in $a$.
      For the Kansas, the dashed line is set at $h=0.57$, Hack's exponent
      estimated via simple regression analysis for points
      with $10^{7} < a < 10^{10}$ m$^2$.
      The local exponent is seen to gradually rise through
      the $h=0.57$ level indicating scaling is not robust.
      For the Mississippi in (b), the dashed line is 
      a Hack exponent of $0.55$ calculated
      from regression on data in the interval $10^{9} < a < 10^{11.5}$ m$^2$.
      The local Hack exponent is seen to
      gradually rise and fall about this value.
      }
    \label{fig:dev.diffmeanhack_kansas}
  \end{center}
\end{figure}

To determine $h$,
we consider Hack's law (equation~\req{eq:dev.hackslaw3})
explicitly in logarithmic coordinates,
\begin{equation}
  \label{eq:hackslawlog}
  \log_{10}{\tavg{l(a)}} = \log_{10}\theta + h\log_{10}{a}.
\end{equation}
The derivative of this equation
with respect to $\log_{10}{a}$ then gives Hack's
exponent as a function of area,
\begin{equation}
  \label{eq:hackslawlog_deriv}
  h(a) = \ddiff{\log_{10}a} \log_{10}{\avg{l(a)}}.
\end{equation}
We may think of $h(a)$ as a ``local Hack exponent.''
Note that non-constant trends in $h(a)$ 
indicate scaling does not hold.
We calculate the discrete derivative
as above for the Kansas and Mississippi.
We smooth the data by taking running
averages with varying window sizes of $n$
samples, the results for $n=32$ being shown in
Figures~\ref{fig:dev.diffmeanhack_kansas}(a)
and~\ref{fig:dev.diffmeanhack_kansas}(b)
where the spacing of $\log{a}$ is $0.02$ orders of magnitude.  
Thus, the running averages for the 
figures are taken over corresponding
area ranges 0.64 orders of magnitude.

Now, if the scaling law in question is truly a scaling law,
the above type of derivative will 
fluctuate around a constant value of exponent
over several orders of magnitude.
With increasing $n$, these fluctuations will
necessarily decrease and we should see
the derivative holding steady around the exponent's value.

At first glance, we notice considerable variation
in $h(a)$ for both data sets with the Kansas
standing out.  Fluctuations are reduced with 
increasing $n$ but we observe continuous variation
of the local Hack exponent with area.
For the example of the Kansas, the linear regime and ensuing crossover
appear as a steep rise followed
by a drop and then another rise during all of which the local
Hack exponent moves well below $1/2$.

It is after these small scale fluctuations
that we would expect to find Hack's exponent.
For the Kansas river data, we see the derivative
gradually climbs for all values of $n$ before reaching
the end of the intermediate regime where
the putative scaling breaks down altogether.

For the Kansas show in
Figure~\ref{fig:dev.diffmeanhack_kansas}(a),
the dashed line represents $h=0.57$, our
estimate of Hack's exponent from simple
regression on the mean-Hack distribution
of Figure~\ref{fig:dev.meanhackdist_kansas}(a).
For the regression calculation,
the intermediate region was identified
from the figure to be $10^{7} < a < 10^{10}$ m$^2$.
We see from the smoothed derivative
in Figure~\ref{fig:dev.diffmeanhack_kansas}(a)
that the value $h=0.57$ is not precise.
After the crossover from
the linear region has been completed,
we observe a slow rise from
$h \simeq 0.54$ to $h \simeq 0.63$.
Thus, the local Hack exponent $h(a)$ gradually
climbs above $h=0.57$ rather than fluctuate
around it.

A similar slow change in $h(a)$ is
observed for the Mississippi data.
We see in Figure~\ref{fig:dev.diffmeanhack_kansas}(b)
a gradual rise and then fall in $h(a)$.
The dashed line here represents $h=0.55$,
the value of which was determined from 
Figure~\ref{fig:dev.meanhackdist_kansas}(b)
using regression on the range
$10^{9} < a < 10^{11.5}$ m$^2$.
The range of $h(a)$ is roughly $[0.52,0.58]$.
Again, while $h=0.55$ approximates the
derivative throughout this intermediate
range of Hack's law, we cannot claim
it to be a precise value.

We observe the same drifts in $h(a)$
in other datasets and for varying window size $n$
of the running average.
The results suggest that we cannot assign
specific Hack exponents
to these river networks
and are therefore unable to even
consider what might be an
appropriate universality class.
The value of $h$ obtained by regression
analysis is clearly sensitive to the
the range of $a$ used.
Furthermore, these results indicate that we 
should maintain healthy reservations about
the exact values of other reported exponents.

\section{Deviations at large scales}
\label{subsec:dev.large}

We turn now to deviations from Hack's law at large scales.
As we move beyond
the intermediate region of approximate scaling, 
fluctuations in $h(a)$ begin to grow rapidly.
This is clear on inspection
of the derivatives of Hack's law in
Figures~\ref{fig:dev.diffmeanhack_kansas}(a)
and~\ref{fig:dev.diffmeanhack_kansas}(b).
There are two main factors conspiring
to drive these fluctuations up.
The first is that the number of samples
of sub-basins with area $a$
decays algebraically in $a$.  This is just
the observation that $P(a) \propto a^{-\tau}$
as per equation~\req{eq:dev.PaPl}.
The second factor is that fluctuations in $l$ and $a$
are on the order of the parameters themselves.
This follows from our generalization of Hack's law
which shows, for example, that the moments 
$\tavg{l^q}$ of $\cprob{l}{a}$ grow like $a^{qh}$.
Thus, the standard deviation grows like the mean:
$\sigma(l) = (\tavg{l^2}-\tavg{l}^2)^{1/2} \propto a^h \propto \tavg{l}$.

\subsection{Stream ordering and Horton's laws}

So as to understand these large scale deviations from Hack's law,
we need to examine network structure in depth.
One way to do this is by using Horton-Strahler 
stream ordering~\cite{horton45,strahler57}
and a generalization
of the well-known Horton's 
laws~\cite{horton45,schumm56a,peckham99,dodds99pa,dodds2000ub}.
This will naturally allow us to deal with the
discrete nature of a network that is most
apparent at large scales.

Stream ordering discretizes
a network into a set of stream segments 
(or, equivalently, a set of nested basins)
by an iterative pruning.
Source streams (i.e., those without tributaries) are
designated as stream segments of order $\om=1$.
These are removed from the network and the new source
streams are then labelled as order $\om=2$ stream segments.
The process is repeated until a single stream segment
of order $\om=\Om$ is left and the basin itself is
defined to be of order $\Om$.

Natural metrics for an ordered river network are 
$n_\om$, the number of order $\om$ stream segments (or basins),
$\bar{a}_\om$, the average area of order $\om$ basins,
$\bar{l}_\om$, the average main stream length of order $\om$ basins,
and
$\okellsombar$, the average length of order $\om$ stream segments.
Horton's laws state that these quantities change 
regularly from order to order, i.e.,
\begin{equation}
  \label{eq:dev.horton}
  \frac{n_{\om}}{n_{\om+1}} = R_n
  \quad\mbox{and}\quad
  \frac{\bar{X}_{\om+1}}{\bar{X}_{\om}} = R_X,
\end{equation}
where $X = a$, $l$ or $\okell$.
Note that all ratios are defined to be greater than unity
since areas and lengths increase but number decreases.
Also, there are only two independent
ratios since $R_a \equiv R_n$ and
$R_l \equiv R_\okell$~\cite{dodds99pa}.
Horton's laws mean that stream-order quantities
change exponentially with order.  For example, 
\req{eq:dev.horton} gives that
$l_\om \propto (R_l)^\om$.

\subsection{Discrete version of Hack's law}

\begin{figure}[tb!]
  \begin{center}
    \ifthenelse{\boolean{@twocolumn}}
    {
    \epsfig{file=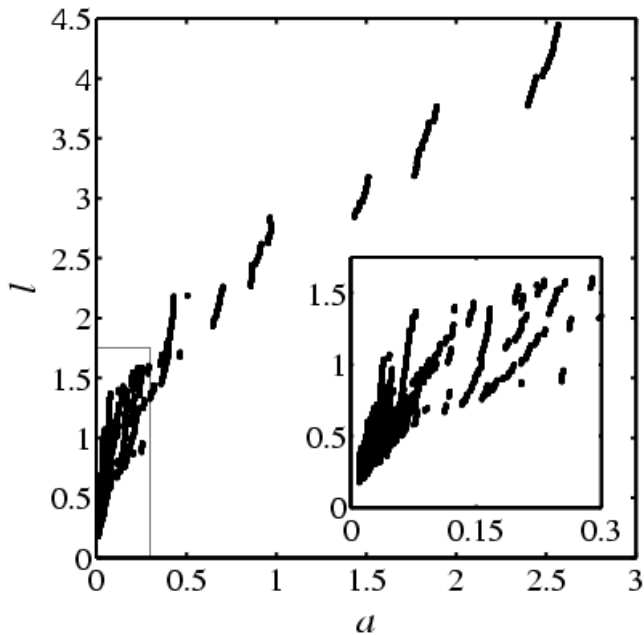,width=0.48\textwidth}
    }
    {
    \epsfig{file=fig_al_ge10e10_mispi10_noname.ps,width=0.48\textwidth}
    }
    \caption[Hack distribution in linear coordinates for the Mississippi]{
      Hack distribution for the Mississippi 
      plotted in linear space.  
      The area units $a$ is $10^{12}$ m$^2$ and 
      length $l$ is $10^{6}$ m.
      The discreteness of the
      basin structure is clearly 
      indicated by the isolated, linear fragments.
      The inset is a blow-up of the box on the main graph.
      }
    \label{fig:dev.al_ge10e10_mispi10}
  \end{center}
\end{figure}

Returning to Hack's law, we examine
its large scale fluctuations with the help of stream ordering.
We are interested in the size of these fluctuations
and also how they might correlate with
the overall shape of a basin.
First, we note that the structure of the network
at large scales is explicitly discrete.
Figure~\ref{fig:dev.al_ge10e10_mispi10}
demonstrates this by plotting the distribution
of $(a,l)$ without the usual logarithmic transformation.
Hack's law is seen to be composed of linear fragments.
As explained above in Figure~\ref{fig:dev.hackjump}, 
areas and length increase
in proportion to each other along streams where
no major tributaries enter.  As soon as a stream
does combine with a comparable one, a
jump in drainage area occurs.  Thus,
we see in Figure~\ref{fig:dev.al_ge10e10_mispi10}
isolated linear segments which upon ending
at a point $(a_1,l_1)$ begin again at $(a_1+a_2,l_1)$,
i.e., the main stream length stays the same but
the area is shifted.

We consider a stream ordering version of Hack's law
given by the points $(\bar{a}_\om,\bar{l}_\om)$.
The scaling of these data points is
equivalent to scaling in the usual Hack's law.
Also, given Horton's laws, it follows that
$h = \ln{R_l}/\ln{R_n}$ (using $R_a \equiv R_n$).
Along the lines of the derivative we
introduced to study intermediate scale
fluctuations in equation~\req{eq:hackslawlog_deriv}, 
we have here an order-based difference:
\begin{equation}
  \label{eq:dev.hackdiff}
  h_{\om,\om-1}
  = \frac{\log{\bar{l}_{\om}/\bar{l}_{\om-1}}}
  {\log{\bar{a}_{\om}/\bar{a}_{\om-1}}}.
\end{equation}
We can further extend this definition
to differences between non-adjacent orders:
\begin{equation}
  \label{eq:dev.hackdiff2}
  h_{\om,\om'}
  = \frac{\log{\bar{l}_{\om}/\bar{l}_{\om'}}}
  {\log{\bar{a}_{\om}/\bar{a}_{\om'}}}.
\end{equation}
This type of difference, 
where $\om' < \om$,
may be best thought of as
a measure of trends rather
than an approximate discrete derivative.

Using these discrete differences,
we examine two features of the order-based
versions of Hack's law.
First we consider correlations between
large scale deviations within an
individual basin and second,
correlations between overall
deviations and basin shape.
For the latter, we will also consider
deviations as they move back
into the intermediate scale.
This will help to explain
the gradual deviations from
scaling we have observed at
intermediate scales.

Since deviations at large scales
are reflective of only a few basins,
we require an ensemble of basins
to provide sufficient statistics.
As an example of such an ensemble,
we take the set of order $\Om=7$ basins
of the Mississippi basin.
For the dataset used here
where the overall basin itself
is of order $\Om=11$,
we have 104 order $\Om=7$ sub-basins.
The Horton averages for
these basins are
$\bar{a}_7 \simeq 16600$ km$^2$,
$\bar{l}_7 \simeq 350$ km,
and $\bar{L}_7 \simeq 210$ km.

\begin{figure}[htb!]
  \begin{center}
    \epsfig{file=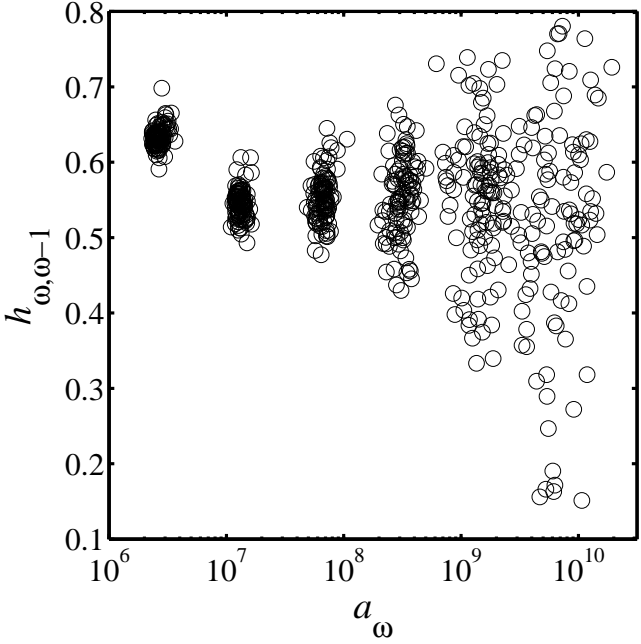,width=0.48\textwidth}
    \caption[Differences for stream order-based version of Hack's law]{
      Differences of the stream order-based version of Hack's law
      for 104 order $\Om=7$ basins of the Mississippi
      (compare the continous versions given in
      Figure~\ref{fig:dev.diffmeanhack_kansas})
      The plots are overlaid to give a sense of the increase
      in fluctuations of the local Hack exponent $h_{\om,\om-1}$
      with increasing order $\om$.  The clusters correspond
      to $\om=1,2,\ldots,7$, moving across from left to right.
      }
    \label{fig:dev.orderhack7_mispi10_2}
  \end{center}
\end{figure}

For each basin, we first calculate the Horton
averages $(\bar{a}_\om,\bar{l}_\om)$.
We then compute $h_{\om,\om-1}$, the Hack difference
given in equation~\req{eq:dev.hackdiff}.
To give a rough picture of what is observed,
Figure~\ref{fig:dev.orderhack7_mispi10_2}
shows a scatter plot
of $h_{\om,\om-1}$ for all order $\Om=7$ basins.
Note the increase in fluctuations with increasing $\om$.
This increase is qualitatively consistent with the smooth versions found
in the single basin examples of 
Figures~\ref{fig:dev.diffmeanhack_kansas}(a)
and~\ref{fig:dev.diffmeanhack_kansas}(b).
In part, less self-averaging for larger $\om$ results
in a greater spread in this discrete derivative.
However, as we will show,
these fluctuations are also correlated with
fluctuations in basin shape.

\subsection{Effect of basin shape on Hack's law}

In what follows, we extract two statistical
measures of correlations between 
deviations in Hack's law and overall basin shape.
These are $r$, the standard linear correlation coefficient
and $r_s$, the 
Spearman rank-order correlation 
coefficient~\cite{press92,lehmann75,sprent93}.
For $N$ observations of data pairs
$(u_i,v_i)$, $r$ is defined to be
\begin{equation}
  \label{eq:dev.rdef}
  r = 
  \frac{\sum_{i=1}^{N}(u_i-\mu_u)(v_i-\mu_v)}
  {\sum_{i=1}^{N}(x_i-\mu_u)^2
    \sum_{i=1}^{N}(y_i-\mu_v)^2}
  = \frac{C(u,v)}
  {\sigma_u \sigma_v},
\end{equation}
where $C(u,v)$ is the covariance of the $u_i$'s and $v_i$'s,
$\mu_u$ and $\mu_v$ their means,
and $\sigma_u$ and $\sigma_v$ their standard deviations.
The value of Spearman's $r_s$ is determined
in the same way but for the $u_i$ and $v_i$
replaced by their ranks.  From $r_s$, we determine
a two-sided significance $p_s$ via
Student's t-distribution~\cite{press92}.

We define $\kappa$, a measure of
basin aspect ratio, as
\begin{equation}
  \label{eq:dev.aspectratio}
  \kappa = L^2/a.
\end{equation}
Long and narrow basins correspond
to $\kappa \gg 1$ while for
short and wide basins, we have $\kappa \ll 1$ 

\begin{figure}[tb!]
  \begin{center}
    \epsfig{file=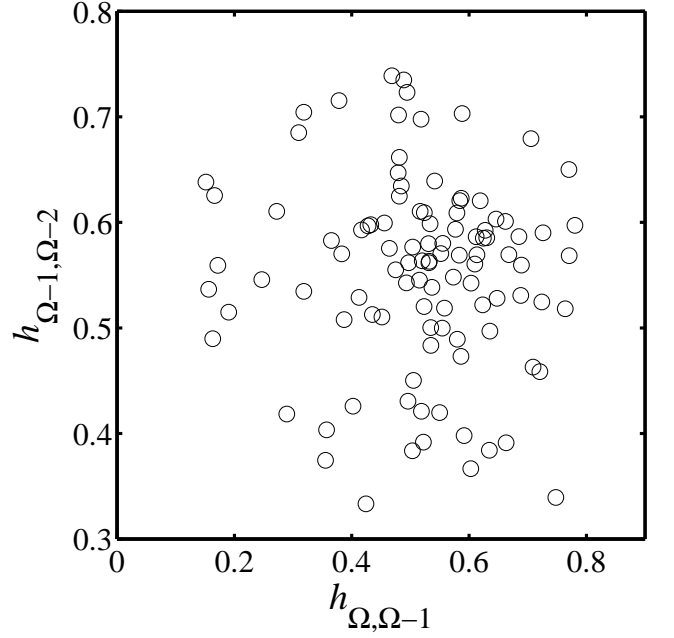,width=0.48\textwidth}
    \caption[Correlations of intra-basin large scale fluctuations in Hack's law]{
      A comparison of the stream order Hack
      derivatives $h_{\Om,\Om-1}$ and $h_{\Om-1,\Om-2}$
      for each of the 104 order $\Om=7$ basins of the Mississippi.
      The linear correlation coefficient is $r=-0.06$
      and the Spearman correlation coefficient is $r_s=-0.08$.
      The latter has probability $p_s=0.43$ indicating
      there are no significant correlations.
      }
    \label{fig:dev.hackorder_correls}
  \end{center}
\end{figure}

We now examine the
discrete derivatives of Hack's law in more detail.
In order to discern correlations
between large scale fluctuations 
within individual basins, 
we specifically look at
the last two differences
in a basin:
$h_{\Om,\Om-1}$ and $h_{\Om-1,\Om-2}$.
For each of the Mississippi's 104 order $\Om=7$ 
basins, these values are plotted
against each other
in Figure~\ref{fig:dev.hackorder_correls}.
Both our correlation measurements
strongly suggest these differences are uncorrelated.
The linear correlation coefficient 
is $r = -0.06 \simeq 0$ and,
similarly, we have $r_s = -0.08 \simeq 0$.
The significance $p_s=0.43$ implies that
the null hypothesis of uncorrelated
data cannot be rejected.

\begin{figure}[tb!]
  \begin{center}
    \epsfig{file=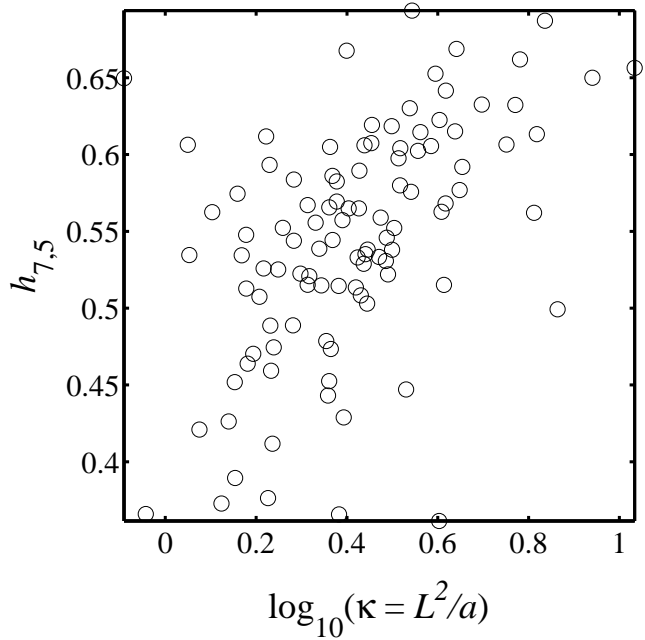,width=0.48\textwidth}
    \caption[Correlation between trends in Hack's law and basin shape]{
      Correlation between trends in Hack's law and
      the aspect ratio of a basin as estimated by
      $\kappa = L^2/a$.  
      The data is for the
      order $\Om=7$ basins of the Mississippi
      and the specific trend is $h_{7,5}$.
      The correlation measurements give
      $r=0.50$, $r_s=0.53$ and
      $p_s < 10^{-8}$.
      }
    \label{fig:dev.orderhack7_asp5_7log_mis}
  \end{center}
\end{figure}

Thus, for Hack's law in an individual basin, large
scale fluctuations are seen to be uncorrelated.
However, correlations between these fluctuations
and other factors may still exist.
This leads us to our second test which concerns
the relationship between trends in Hack's law and
overall basin shape.

Figure~\ref{fig:dev.orderhack7_asp5_7log_mis}
shows a comparison of the aspect ratio $\kappa$
and $h_{7,5}$ for the order $\Om=7$
basins of the Mississippi.
The measured correlation coefficients are
$r=0.50$ and $r_s=0.53$,
giving a significance of $p_s < 10^{-8}$.
Furthermore, we find the differences $h_{7,6}$
($r=0.34$, $r_s=0.39$ and $p_s < 10^{-4}$)
and $h_{6,5}$ 
($r=0.35$, $r_s=0.34$ and $p_s < 10^{-3}$)
are individually correlated
with basin shape.  
We observe this correlation between basin shape 
and trends in Hack's law at
large scales, namely $h_{\Om,\Om-1}$, $h_{\Om-1,\Om-2}$
and $h_{\Om,\Om-2}$, repeatedly
in our other data sets.  
In some cases, correlations
extend further to $h_{\Om-2,\Om-3}$.

Since the area ratio $R_a$ is typically 
in the range 4--5,  Hack's law
is affected by boundary conditions set by
the geometry of the overall basin
down to sub-basins one 
to two orders of magnitude
smaller in area than the overall basin.
These deviations are present
regardless of the absolute size of the overall basin.
Furthermore,
the origin of the basin boundaries being geologic or 
chance or both is irrelevant---large scale deviations will still occur.
However, it is reasonable to suggest
that particularly strong deviations
are more likely the result of geologic structure rather than
simple fluctuations.

\section{Conclusion}
\label{sec:dev.conclusion}

Hack's law is a central
relation in the study of river networks
and branching networks in general.
We have shown Hack's law to have 
a more complicated structure than
is typically given attention.
The starting generalization is
to consider fluctuations around
scaling.  Using the directed, random network
model, a form for the Hack distribution underlying
Hack's law may be postulated
and reasonable agreement with real networks
is observed.  
Questions of the validity of the distribution
aside, the Hack mean coefficient $\theta$ and
the Hack standard deviation coefficient $\eta$
should be standard measurements because
they provide further points of comparison
between theory and other basins.

With the idealized Hack distribution proposed,
we may begin to understand deviations from its form.
As with any scaling law pertaining
to a physical system, cutoffs
in scaling must exist and need to be understood.
For small scales, we have identified 
the presence of
linear sub-basins as the source
of an initial linear relation between
area and stream length.
At large scales, statistical fluctuations and
geologic boundaries give rise to basins whose overall
shape produces deviations in Hack's laws.
Both deviations extend over a considerable
range of areas as do the crossovers which
link them to the region of intermediate scales,
particularly the crossover from small scales.

Finally, by focusing in detail on a few
large-scale examples networks, 
we have found evidence
that river networks do not 
belong to well defined universality classes.
The relationship between basin area and 
stream length may be approximately,
and in some cases very well, described
by scaling laws but not exactly so.
The gradual drift in exponents we observe
suggests a more complicated picture,
one where subtle correlations between basin shape
and geologic features are 
intrinsic to river network structure.

\section*{Acknowledgements}
This work was supported in part by NSF grant EAR-9706220
and the Department of Energy grant DE FG02-99ER 15004.


\begin{thebibliography}{10}
\bibitem{note:dev.nilecongo}
\bibinfo{note}{The data for the Nile and Congo was obtained from the United
  States Geological Survey's 30-arc-second Hydro1K dataset which may be
  accessed on the Internet at \www{edcftp.cr.usgs.gov}. Note that these Hydro1K
  datasets have undergone the extra processing of projection onto a uniform
  grid.}
\end{thebibliography}
\end{document}